\documentclass[11pt]{article}
\usepackage[affil-it]{authblk}
\usepackage{amsmath,amssymb}
\usepackage{color}
\usepackage{graphicx}
\usepackage[top=2cm, bottom=2cm, left=2cm, right=2cm]{geometry}

\makeatletter
\newenvironment{highlightequation}
{\def\tagform@##1{\maketag@@@{(\ignorespaces##1\unskip\@@italiccorr*)}}%
\ignorespaces}
{\def\tagform@##1{\maketag@@@{(\ignorespaces##1\unskip\@@italiccorr)}}%
\ignorespacesafterend}
\makeatother

\newcommand{\II}{I\!\!I}
\newcommand{\III}{I\!\!I\!\!I}
\DeclareMathOperator{\tr}{tr}

\begin{document}

\title{A non-linear rod model for folded elastic strips}

\author{Marcelo A. Dias \thanks{marcelo\_dias@brown.edu}}
\affil{School of Engineering, Brown University, Providence, Rhode Island 02912, USA}  

\author{Basile Audoly\thanks{audoly@lmm.jussieu.fr}}
\affil{UPMC Univ Paris 06,CNRS, UMR 7190, Institut Jean Le Rond d'Alembert, F-75005 Paris, France}

\date{\today}
\maketitle

\begin{abstract}
We consider the equilibrium shapes of a thin, annular strip cut out in
an elastic sheet.  When a central fold is formed by creasing beyond
the elastic limit, the strip has been observed to buckle out-of-plane.
Starting from the theory of elastic plates, we derive a Kirchhoff rod
model for the folded strip.  A non-linear effective constitutive law
incorporating the underlying geometrical constraints is derived, in
which the angle the ridge appears as an internal degree of freedom.
By contrast with traditional thin-walled beam models, this
constitutive law captures large, non-rigid deformations of the
cross-sections, including finite variations of the dihedral angle at
the ridge.  Using this effective rod theory, we identify a buckling
instability that produces the out-of-plane configurations of the
folded strip, and show that the strip behaves as an elastic ring
having one frozen mode of curvature.  In addition, we point out two
novel buckling patterns: one where the centerline remains planar and
the ridge angle is modulated; another one where the bending
deformation is localized.  These patterns are observed experimentally,
explained based on stability analyses, and reproduced in simulations
of the post-buckled configurations.
\end{abstract}

\section{Introduction}
\begin{figure}
    \centering
    \includegraphics[width=.99\textwidth]{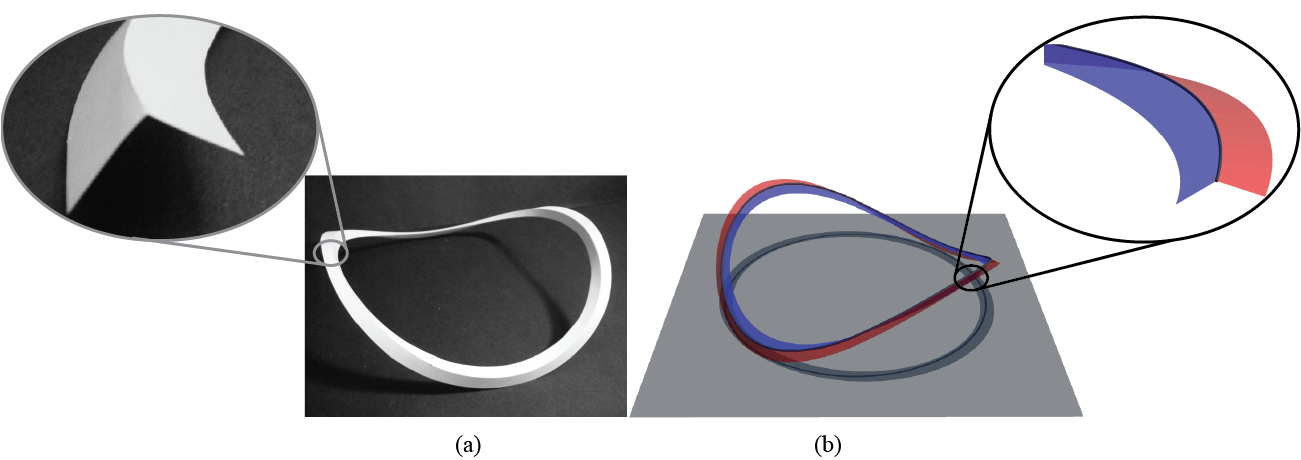}
    \caption{Buckling of an annular elastic strip having a central
    fold: (a) model cut out in an initially flat piece of paper; (b)
    one of the goals of this paper is to represent this folded strip
    as a thin elastic rod, the ridge angle being considered as an
    internal degree of freedom.}
    \label{fig:RodSection}
\end{figure}

Although the idea that a sheet of paper can be folded along an
arbitrary curve is unfamiliar to many, performing this activity has
been a form of art for quite some time.  Bauhaus, the extinct German
school of art and design, was a pioneer in developing the concept of
curved folding structures by the end of the 1920s \cite{Wingler69}.
This practice often yields severely buckled and mechanically stiff
sculptures featuring interesting structural properties and reveals
new ways to think about engineering and architecture \cite{Engel1997,Jackson2011,Schenk2011}. 
Traditional origami has had a strong influence in the solution of many practical problems, 
to cite a few, the deployment of large membranes in space \cite{Miura1980} and biomedical applications \cite{Kuribayashi2006}. However, exploring this long established art form still has a lot of potential.  
Since the work by Huffman in 1976 \cite{Huffman76}, an
elegant and groundbreaking description of the geometry of curved
creases, more attention has been devoted to this subject
\cite{DuncanDuncan82,FuchsTabachnikov99,PottmannWallner2001,
KilianFloeryChen-Curved-folding-2008}.  A mechanical approach of
structures comprising curved creases has recently been
proposed~\cite{Dias2012a} motivated by the intriguing 3d shapes shown
in figure~\ref{fig:RodSection}.  In the present paper, we build upon
this recent work by further exploring the mechanical models governing
folded structures.

Folded structures combine geometry and mechanics: they deform in an
inextensible manner and their mechanics is constrained by the geometry
of developable
surfaces~\cite{Spivak-A-comprehensive-introduction-to-differential-geometry-1979,doCarmo76}.
Here, we consider one of the simplest folded structures: a narrow
elastic plate comprising a central fold, as shown in
figure~\ref{fig:RodSection}.  The role of geometry is apparent from
the following observations, which anyone can reproduce with a paper
model: the curvature of the crease line is minimum when the fold is
flattened, a closed crease pattern results into a fold that buckles
out of plane, while an open crease pattern results in a planar fold.
These and other geometrical facts have been proved by Fuchs and
Tabachnikov~\cite{FuchsTabachnikov99}.

The mechanics of thin rods has a long
history~\cite{Dill1992,Love,Antman1995}, and is used to
tackle a number of problems from different fields today, such as the
morphogenesis of slender objects~\cite{Wolgemuth2004266,Moulton2013},
the equilibrium shape of elongated biological filaments --- such as
DNA~\cite{Shi1994} and bacterial flagellum~\cite{Powers2010} --- and
the mechanics of
the human hair~\cite{Audoly-Pomeau-Elasticity-and-geometry:-from-2010,%
Goldstein-Warren-EtAl-Shape-of-a-Ponytail-and-the-Statistical-2012}.
The classical theory of rods, known as Kirchhoff's rod theory, assumes
that all dimensions of the cross-section are comparable: the
consequence is that the cross-sections of the rod deform almost
rigidly as long as long as the strain remains small.  This assumption
does not apply to a folded strip: its cross-sections are slender, as
shown in the inset of figure~\ref{fig:RodSection}(a), and, as a result,
they can bend by a large amount.  In addition, the dihedral angle at
the ridge can also vary by a large amount.

Vlasov's theory for thin-walled beams overcomes the limitations of
Kirchhoff's theory by relaxing some kinematic constraints and
considering additional modes of deformations of the cross-section.
This kinematic enrichment can be justified from 3d elasticity:
assuming a thin-walled geometry, asymptotic convergence of the 3d
problem to a rod model of Vlasov type has been
established formally~\cite{Hamdouni-Millet-An-asymptotic-non-linear-model-2006,%
Hamdouni-Millet-An-asymptotic-linear-thin-walled-2011}.
This justification from 3d elasticity requires that the deformations 
are mild, however: the cross-sections can only bend by a small amount 
away from their natural shape.

Mechanical models have been proposed to capture the large deformations
of thin-walled beams.  The special case of curved cross-sections must
be addressed starting from the theory of shell: in this case, the bending of
the centerline involves a trade-off between the shell's bending and
stretching
energies~\cite{Mansfield-Large-Deflexion-Torsion-and-Flexure-1973,%
Seffen2000,Guinot2012,%
Giomi-Mahadevan-Multi-stability-of-free-spontaneously-2012}.  By
contrast, the strip that we consider is developable; it can be
studied based on an inextensible plate model, in which the stretching
energy plays no role.  A model for a thin elastic strip has been
developed Sadowsky~\cite{Sadowsky30} in the case of a narrow ribbon,
and later extended by Wunderlich to a finite
width~\cite{Wunderlich-Uber-ein-abwickelbares-Mobiusband-1962}.  These
strip models have found numerous applications recently,
see~\cite{StarostinHeijden08} for instance.  They have been developed
independently of the theory of rods, as they make use of unknowns that
are tied to the developability constraint. 

Here, we develop a unified view of strips and rods.  We promote the
viewpoint that elastic strips are just a special case of thin rods.
The equations for the equilibrium of a narrow, inextensible plate are
shown to be equivalent to those for an inextensible rod.  To establish
this equivalence, we identify the relevant geometrical
constraints and derive of an effective, non-linear constitutive law.
By doing so, we extend Vlasov's models for thin-walled beams to large
deformations.  A unified perspective of strips and rods brings in the
following benefits: instead of re-deriving the equations of
equilibrium for strips from scratch, which is cumbersome, we show that
the classical Kirchhoff equations are applicable; we identify for the
first time the stress variables relevant to the strip model, which is
crucial for stability problems; the extension of the strip model to
handle natural (geodesic) curvature, or the presence of a central fold
becomes straightforward, as we demonstrate; stability analyses and
numerical solutions of post-buckled equilibria can be carried out in
close analogy with what is routinely done for classical rods.


This paper is organized as follows.  In section~\ref{sec:singleStrip},
we start by the smooth case, \emph{i.e.}\ consider an elastic strip
without a fold, and derive an equivalent rod model for it.  In
section~\ref{sec:bistrip}, we extend this model to a folded strip,
which we call a bistrip; this is one of the main results of our paper.
In section~\ref{sec:circular}, we derive circular solutions for the
bistrip.  Their stability is analyzed in
section~\ref{sec:linearStability}, and we identify two families of
buckling modes: one family of modes explains the typical non-planar
shapes of the closed bistrip reported earlier, while the second mode
of buckling is novel.  The predictions of the linear stability
analysis are confronted to experiments in
section~\ref{sec:experiments}, and to simulations of the post-buckled
solutions in section~\ref{sec:AUTO}.

\section{Smooth case: equivalent rod model for a curved elastic strip}
\label{sec:singleStrip}

We start by considering the case of a narrow strip having no central
fold and show how it can be described using the language of thin
elastic rods. The model we derive extends the model of
Sadowsky~\cite{Sadowsky30} to account for the geodesic curvature of
the strip and bridges the gap between his formulation and the
classical theory of elastic rods.

\subsection{Kinematics and constraints}

\begin{figure}
    \centering
    \includegraphics{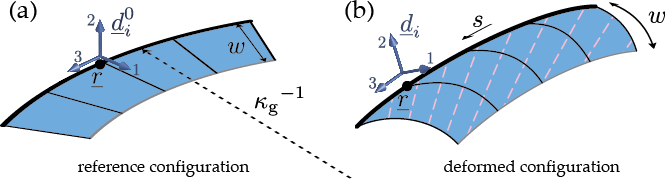}
    \caption{Analysis of a narrow elastic strip, without any fold.
    (a) One of the lateral boundaries is used as a centerline (thick
    curve), and its curvature in the reference configuration defines
    the geodesic curvature $\kappa_{\mathrm{g}}$.  (b) The underlying
    mechanical model is an inextensible plate: in the deformed
    configuration, the generatrices, shown by the dashed lines, can
    make an arbitrary angle with the tangent $\underline{d}_{3}$ to
    the centerline.  As a result, the cross-section (thin solid lines)
    can be significantly curved.  The inextensibility constraint is
    used to reconstruct the mid-surface of the plate, based on the
    centerline shape $\underline{r}(s)$ (thick curve) and on the
    material frame $\underline{d}_{i}(s)$: this allows the strip to
    be viewed as a thin rod.}
    \label{fig:singleStrip}
\end{figure}

We consider an inextensible elastic plate of thickness $h$ and width
$w$ with a large aspect-ratio, $h\ll w$.  In its undeformed
configuration, the strip is planar.  Under the action of a mechanical
load, it is deformed into a 3d shape, as sketched in
figure~\ref{fig:singleStrip}.
The deformed strip is parameterized by a space curve
$\underline{r}(s)$, called the \emph{centerline}, and by an
orthonormal frame $(\underline{d}_{1}(s), \underline{d}_{2}(s),
\underline{d}_{3}(s))$, called the \emph{material frame}.  To define
the centerline, we pick one of the lateral edges, which is the thick
curve in the figure~\ref{fig:singleStrip}.  The name `centerline' is used for consistency with the theory of rods even though this curve is off the center of
the strip.  We denote by $s$ be the arc-length along this edge, and by
$\underline{r}(s)$ the position in space of the centerline.  The
derivation with respect to arc-length is denoted by a prime.  By
definition, the tangent $\underline{r}'(s)$ is a unit vector.  Since
the plate is inextensible, the arc-length $s$ will be used as a
Lagrangian variable.  The direct orthonormal frame
$\underline{d}_{i}(s)$, with $i=1,2,3$, is defined in such a way that
$\underline{d}_{3}$ is the tangent to the centerline,
\begin{equation}
    \underline{d}_{3}(s) = \underline{r}'(s)
    \label{eq:tangent}
\end{equation}
and that $\underline{d}_{1}(s)$ and $\underline{d}_{3}(s)$ span the
tangent plane to the midsurface of the elastic strip at the point
$\underline{r}(s)$.  Then, the unit vector $\underline{d}_{2}(s) =
\underline{d}_{3}(s)\times \underline{d}_{1}(s)$ is normal to the
midsurface at the edge $\underline{r}(s)$. By construction, the 
material frame is orthonormal and direct:
\begin{equation}
    \underline{d}_{i}(s)\cdot \underline{d}_{j}(s) = \delta_{ij}
    \textrm{,}
    \label{eq:orthonormal}
\end{equation}
for any indices $i,j=\{1,2,3\}$ and for any $s$.  Here, $\delta_{ij}$
denotes Kronecker's symbol, equal to $1$ if $i=j$ and $0$ otherwise. 


The rate of rotation of the material frame with respect to the
arc-length is captured by a vector $\underline{\omega}(s)$, which we
call the Darboux vector or the twist-curvature strain.  It is such
that, for any $i=1,2,3$,
\begin{equation}
    \underline{d}_{i}'(s) = \underline{\omega}(s)\times 
    \underline{d}_{i}(s)
    \textrm{.}
    \label{eq:darbouxVectorDefinition}
\end{equation}
In classical rod theories, the rotation gradient
$\underline{\omega}(s)$ measures the strain associated with the
bending and twisting modes.  Here, we use a plate model, and the
bending strain is measured by the curvature form (second fundamental
form) of the mid-surface, which we denote by
$\underline{\underline{k}}$.  Near a generic point $\underline{r}(s)$
on the centerline, we use the frame $(\underline{d}_{3}(s) ,
\underline{d}_{1}(s))$ tangent to surface: in this frame, the
curvature tensor $\underline{\underline{k}}(s)$ is represented by a
symmetric matrix,
\begin{equation}
    \underline{\underline{k}}(s) =
    \begin{pmatrix}    k_{33}(s) & k_{13}(s) \\
	k_{13}(s) & k_{11}(s)
    \end{pmatrix}_{(\underline{d}_{3},\underline{d}_{1})}
    \textrm{.}
    \nonumber
\end{equation}
From the differential geometry of
surfaces~\cite{Spivak-A-comprehensive-introduction-to-differential-geometry-1979,doCarmo76},
the gradient of the unit normal to a surface along any tangent
direction can be computed from the second fundamental form.  In
particular, if we consider the gradient of the normal
$\underline{d}_{2}(s)$ along the tangent $\underline{d}_{3}(s)$ to the
centerline, we have $\underline{d}_{2}'(s) =
-\underline{\underline{k}}(s)\cdot \underline{d}_{3}(s) =
-(k_{13}(s)\,\underline{d}_{1}(s) + k_{33}(s)\,\underline{d}_{3}(s))$.
Identifying with the case $i=2$
equation~(\ref{eq:darbouxVectorDefinition}), we find that the plate's
bending strain and the equivalent rod's curvature strain are related
by: $k_{13}(s) = \omega_{3}(s)$ and $k_{33}(s) = -\omega_{1}(s)$,
where $\omega_{j} = \underline{\omega}\cdot \underline{d}_{j}$ denote
the components of the Darboux vector in the material frame.  We use
this to express the second fundamental form of the midsurface of the
plate, in terms of the Darboux vector $\underline{\omega}$ of the
equivalent rod:
\begin{equation}
    \underline{\underline{k}}(s) = \left(
    \begin{array}{cc}
        -\omega_{1}(s) & \omega_{3}(s)  \\
        \omega_{3}(s)  & k_{11}(s)
    \end{array}
    \right)_{(\underline{d}_{3},\underline{d}_{1})}
    \textrm{.}
    \label{eq:kFromOmega}
\end{equation}

We assume that the midsurface of the plate is inextensible.  This has
two consequences.  First, by Gauss' theorema
egregium~\cite{Spivak-A-comprehensive-introduction-to-differential-geometry-1979},
its Gauss curvature, defined as the determinant of 
$\underline{\underline{k}}$, is zero:
\begin{equation}
    \mathcal{C}_{\mathrm{d}}(\underline{\omega}, k_{11}) = 0,\quad
    \textrm{where }
    \mathcal{C}_{\mathrm{d}}(\underline{\omega}, k_{11}) = - \det 
    \underline{\underline{k}} =
    \omega_{1}\,k_{11}
    +(\omega_{3})^{2} \textrm{.}
    \label{eq:GaussCondition}
\end{equation}
Second, we note that the quantity $\omega_{2}$, which defines the
geodesic curvature of the centerline with respect to the midsurface,
is conserved by
isometries~\cite{Spivak-A-comprehensive-introduction-to-differential-geometry-1979}.
Let $\kappa_{\mathrm{g}} = \omega_{2}^0$ denote the signed curvature
of the edge (centerline) in the flat configuration of reference: in
the actual configuration, the conservation of the geodesic curvature
implies
\begin{equation}
    \mathcal{C}_{\mathrm{g}}(\underline{\omega}) = 0,\quad
    \textrm{where }\mathcal{C}_{\mathrm{g}}(\underline{\omega})
    = \kappa_{\mathrm{g}} - \omega_{2}
    \textrm{.}
    \label{eq:geodesicConstraint}
\end{equation}

In equation~(\ref{eq:kFromOmega}), we have expressed the `microscopic'
strain $\underline{\underline{k}}$ in terms of the strain
$\underline{\omega}$ of the equivalent rod, and of an additional
`internal' strain variable $k_{11}$.
Equation~(\ref{eq:GaussCondition}) is a kinematic constraint.  It
could be used to eliminate the strain variable $k_{11}$ in favor of
$\underline{\omega}$; we will refrain to do so, however, as this
requires the additional assumption $\omega_{1} \neq 0$.
Equation~(\ref{eq:geodesicConstraint}) is a second kinematic
constraint applicable to the equivalent rod model.

\subsection{Sadowsky's elastic energy}

Let us denote by $k^*$ the typical curvature of the plate,
$|\underline{\underline{k}}| \sim k^*$.  We assume that the strip is
narrow, in the sense that the variations of the curvature tensor on
distances comparable to the width $w$ remain small compared to $k^*$:
$w\,|\underline{\nabla}\underline{\underline{k}}| \ll k^*$.
Therefore, we ignore the dependence of the curvature strain on the
transverse coordinate: in the entire cross-section containing the
centerline point $\underline{r}(s)$, we approximate the curvature
tensor by $\underline{\underline{k}}(s)$.  This approximation has been
used in the past to describe plates with straight
centerlines~\cite{Sadowsky30}.  It is possible to go beyond this
approximation~\cite{Wunderlich-Uber-ein-abwickelbares-Mobiusband-1962},
as required when the curvature becomes large at localized spots along
one of the edges~\cite{StarostinHeijden07} --- we will refer to this
model as a strip of finite width.  An even more general model,
applicable to strips of finite width and non-zero geodesic curvature,
has been recently derived in~\cite{Dias2012a}.

We return to our small width approximation: the bending energy of the
inextensible plate $E_{\mathrm{p}}$ can be integrated along the
transverse direction, which yields
\begin{equation}
    E_{\mathrm{p}} = 
    \int \frac{w\,D}{2}\,\left(
    (1-\nu)\,\tr(\underline{\underline{k}}^2)
    +\nu\,\tr^2\underline{\underline{k}}
    \right)
    \,\mathrm{d}s,
    \nonumber
\end{equation}
where $D=E\,h^3/(12(1-\nu^2)))$ is the bending modulus of the plate,
$E$ its Young's modulus, $\nu$ its Poisson's ratio and $h$ its
thickness.  For a $2\times2$ matrix, the following identity holds:
$\tr^2\underline{\underline{k}}
=\tr\left(\underline{\underline{k}}^2\right)+2\,\det\underline{\underline{k}}$.
Dropping the determinant using the inextensibility
condition~(\ref{eq:GaussCondition}), one can rewrite the elastic 
energy as
\begin{equation}
    E_{\mathrm{p}} 
    =\int\frac{B}{2}\,\underline{\underline{k}}:\underline{\underline{k}}
    \,\mathrm{d}s,
    \label{eq:bendingEnergy2}
\end{equation}
where $B = w\,D$ is a rod-type bending modulus, and the double
contraction operator is defined by
$\underline{\underline{a}}:\underline{\underline{b}} =
\tr(\underline{\underline{a}}\cdot \underline{\underline{b}}) =
\sum_{ij}a_{ij}\,b_{ji}$.  Note that when $k_{11}$ is eliminated using
the constraint~(\ref{eq:GaussCondition}), the elastic energy
$E_{\mathrm{p}}$ coincides with that derived by
Sadowsky~\cite{Sadowsky30}:
\begin{equation}
    E_{\mathrm{p}} = 
 \frac{B}{2}   \int\,\left(
    {k_{33}}^2 + {k_{11}}^2 + 2\,{k_{13}}^2
    \right)\,\mathrm{d}s
    =
    \frac{B}{2}\int\,\left(
    {\omega_{1}}^2 
    + 2\,{\omega_{3}}^2
    + \frac{{\omega_{3}}^4}{{\omega_{1}}^2}
    \right)\,\mathrm{d}s
 \label{eq:bendingEnergy3}
\end{equation}

\subsection{Constitutive law}

We derive the equivalent rod model for our thin strip simply by
viewing the energy $E_{\mathrm{p}}$ in
equation~(\ref{eq:bendingEnergy3}) as the energy of a thin rod.  The
equivalent thin rod has one internal degree of freedom $k_{11}$ and is
subjected to two kinematical constraints
$\mathcal{C}_{\mathrm{d}}(\omega_{1},\omega_{3},k_{11})$ and
$\mathcal{C}_{\mathrm{g}}(\omega_{2})$.  In the \ref{sec:VirtualWork},
we derive the equations for a rod of this type.  The condition of
equilibrium of the internal variable reads
\begin{subequations}
    \label{eq:constrainedRodsRestate}
    \begin{equation}
        -\frac{\delta E_{\mathrm{p}}}{\delta k_{11}}
        \label{eq:constrainedRodsRestate-K}
	+ 
	B\,\lambda_{\mathrm{d}}\,\frac{\partial \mathcal{C}_{\mathrm{d}}}{
	\partial k_{11}} 
	+ 
	B\,\lambda_{\mathrm{g}}\,\frac{\partial \mathcal{C}_{\mathrm{g}}}{
	\partial k_{11}}
	= 0
	\textrm{,}
    \end{equation}
    and the constitutive law as
    \begin{equation}
        \underline{m} = \sum_{i=1}^{3}\left(
	\frac{\delta E_{\mathrm{p}}}{\delta \omega_{i}}
	 - B\,
	\lambda_{\mathrm{d}}\,
	\frac{\partial \mathcal{C}_{\mathrm{d}}}{\partial \omega_{i}}
	- B\,
	\lambda_{\mathrm{g}}\,
	\frac{\partial \mathcal{C}_{\mathrm{g}}}{\partial \omega_{i}}
	\right)\,\underline{d}_{i}
        \label{eq:constrainedRodsRestate-ConstitutiveLaw}
	\textrm{.}
    \end{equation}
\end{subequations}
Here, $\delta E_{\mathrm{p}} / \delta \omega_{i}$ and $\delta
E_{\mathrm{p}} / \delta k_{11}$ denote the functional derivative of
the elastic energy $E_{\mathrm{p}}$ with respect to the local strain
$\omega_{i}(s)$ and internal variable $k_{11}(s)$, respectively.
These equations were obtained by extending
equations~(\ref{eq:constitutiveLawForConstrainedRod-Constraint}--\ref{eq:constitutiveLawForConstrainedRod-M})
of the appendix to the case of two constraints, and by identifying the
internal degree of freedom $k=k_{11}$ and the energy $E_{\mathrm{el}}
= E_{\mathrm{p}}$.  The two Lagrange multipliers
$\lambda_{\mathrm{d}}$ and $\lambda_{\mathrm{g}}$ are associated with
the two constraint.  For convenience, they have been rescaled with the
bending modulus $B$, \emph{i.e.}\ the quantity $\lambda$ in the
appendix is replaced with $B\,\lambda_{\mathrm{d}}$ and
$B\,\lambda_{\mathrm{g}}$.

Equation~(\ref{eq:constrainedRodsRestate-K}) can be interpreted as the
cancellation of the total generalized force acting on the internal
variable, which is the sum of the standard force in the first term,
and of constraint
forces~\cite{Lanczos-The-Variational-Principles-of-Mechanics-1970} in
the last two terms.  Similarly, the constitutive law in
equation~(\ref{eq:constrainedRodsRestate-ConstitutiveLaw}) is made up
of the usual contribution in rod theory, whereby the internal moment
is the gradient of the elastic energy with respect to the twist and
curvature strains, plus two other terms which are known as
(generalized) constraint forces.

Using the explicit form of the energy $E_{\mathrm{p}}$ and of the 
constraints $\mathcal{C}_{\mathrm{d}}$ and 
$\mathcal{C}_{\mathrm{g}}$, 
equation~(\ref{eq:constrainedRodsRestate-K}) yields
\begin{equation}
    k_{11} = \lambda_{\mathrm{d}}\,\omega_{1} 
     \label{eq:knSol}
    \textrm{.}
\end{equation}
This equation will be used to eliminate the internal variable $k_{11}$
whenever it appears.  In particular, the developability constraint in
equation~(\ref{eq:GaussCondition}) takes the
form
\begin{equation}
    \tilde{\mathcal{C}}_{\mathrm{d}}(\underline{\omega},\lambda_{\mathrm{d}}) = 0,\quad
    \textrm{where }
    \tilde{\mathcal{C}}_{\mathrm{d}}(\underline{\omega},\lambda_{\mathrm{d}}) = ({\omega_{1}})^2\,\lambda_{\mathrm{d}} + (\omega_{3})^2
    \label{eq:developabilityConstraint-KEliminiated}
    \textrm{.}
\end{equation}

The second 
equation~(\ref{eq:constrainedRodsRestate-ConstitutiveLaw}) yields
\begin{equation}
    \underline{m} = B\,(\omega_{1}\,\underline{d}_{1} +
    2\,\omega_{3}\,\underline{d}_{3})
    - B\,\left(
    \lambda_{\mathrm{d}}\,(k_{11}\,\underline{d}_{1}+2\,\omega_{3}\,\underline{d}_{3})
    - \lambda_{\mathrm{g}}\,\underline{d}_{2}
    \right)
    \textrm{.}
    \nonumber
\end{equation}
Eliminating $k_{11}$ and projecting onto the material basis, we find
the expressions of the twisting and bending moments $m_{i} =
\underline{m}\cdot \underline{d}_{i}$:
\begin{subequations}
    \label{eq:constitutiveReduced}
    \begin{align}
	\label{eq:constitutiveReduced1}
	m_1 &= B\,\left(1-{\lambda}_{\mathrm{d}}^2\right)\,\omega_1
	\\
	\label{eq:constitutiveReduced2}
	m_2 &= B\,\lambda_{\mathrm{g}}\\ 
	\label{eq:constitutiveReduced3}
	m_3 &= 2\,B\,\left(1-\lambda_{\mathrm{d}}\right)\,\omega_{3}
	\textrm{.}
    \end{align}
\end{subequations}
This is the constitutive law for a narrow elastic developable strip in
the language of rods.

%
%

\subsection{Remarks}

The expression of the constraints in
equations~(\ref{eq:geodesicConstraint})
and~(\ref{eq:developabilityConstraint-KEliminiated}) and the
constitutive law~(\ref{eq:constitutiveReduced}) are the main results
of of section~\ref{sec:singleStrip}.  They translate the inextensible
strip model into the language of Kirchhoff's rods.  Even though the
strip is linearly elastic, its constitutive law is effectively
non-linear because of the developability constraint.

We emphasize that the cross-sections of the strip are allowed to bend,
as sketched in figure~\ref{fig:singleStrip}(b): this is required to
preserve developabitility when the centerline is both twisted and
bent, \emph{i.e.} when both $\omega_{1}$ and $\omega_{3}$ are
non-zero.  We found that the strip is \emph{equivalent} to an
Euler-Bernoulli rod --- for classical rods, the latter model is
usually justified by assuming that the cross-sections do not bend but
this is just a coincidence.

The strong formulation of the equilibrium of elastic rods is known as
the Kirchhoff
equations~\cite{Audoly-Pomeau-Elasticity-and-geometry:-from-2010}.  It
can be derived by integration by parts from the principle of virtual 
work, as explained in~\ref{sec:VirtualWork}:
\begin{subequations}
    \label{eq:classicalKirchhoffEquilibrium}
    \begin{align}	
        \underline{n}'(s) + \underline{p}(s) &  = \underline{0}
	\label{eq:classicalKirchhoffEquilibrium-r}
	\\
        \underline{m}'(s) + \underline{r}'(s)\times \underline{n}(s) + 
	\underline{q}(s) & =
	\underline{0} 
	\textrm{.}
	\label{eq:classicalKirchhoffEquilibrium-m}
    \end{align}
\end{subequations}
Here, $\underline{n}%
$ is the
Lagrange multiplier associated with the inextensibility and
Euler-Bernoulli constraints, which can be interpreted as the internal
force.
Given a distribution of external force $\underline{p}(s)$ and moment
$\underline{q}(s)$, one can find the equilibria of the strip by
solving the kinematical
equations~(\ref{eq:tangent}--\ref{eq:darbouxVectorDefinition}), the
contraints~(\ref{eq:geodesicConstraint})
and~(\ref{eq:developabilityConstraint-KEliminiated}),
the constitutive law~(\ref{eq:constitutiveReduced}) and the 
equilibrium~(\ref{eq:classicalKirchhoffEquilibrium})
for the unknowns $\underline{r}(s)$, $\underline{d}_{i}(s)$,
$\omega_{i}(s)$, $\lambda_{\mathrm{d}}(s)$ and
$\lambda_{\mathrm{g}}(s)$.

\section{An equivalent rod model for the bistrip}
\label{sec:bistrip}

We now consider the case of an elastic strip having a central fold, as
shown in figure~\ref{fig:fold1}.  The central fold is represented by
an elastic hinge.  The flaps on both sides of the fold are represented
using the mechanical model derived in section~\ref{sec:singleStrip}.
We call \emph{bistrip} this composite object, made up of the two flaps
and the central fold.  In this section, we derive an equivalent rod
model for the bistrip, which takes into account both the bending
stiffness of the inextensible flaps, and the stiffness of the central
ridge.

\subsection{Kinematics}

\begin{figure}[tbp]
    \centerline{\includegraphics[width=.35\textwidth]{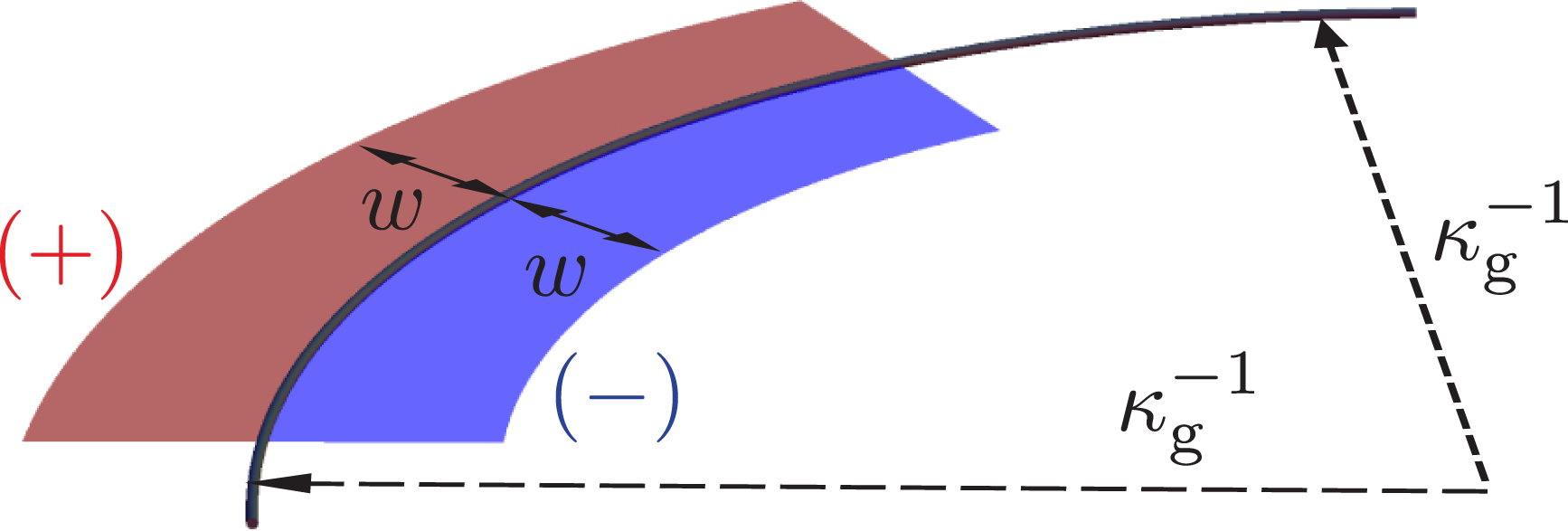}}
    \caption{A strip with a central fold in its planar configuration,
    which we call a bistrip.  The outer and inner flaps are labelled
    by $(+)$ and $(-)$, respectively.  In this planar configuration,
    the curvature of the fold coincides with the geodesic curvature
    $\kappa_{\mathrm{g}}$.}
    \protect\label{fig:FlatBistrip}
\end{figure}

The planar configuration of the bistrip is show in
figure~\ref{fig:FlatBistrip}: in this configuration, the curvature of
the central fold coincides with the geodesic curvature
$\kappa_{\mathrm{g}}$.
The outer and inner flaps, on each side of the central fold, are
labelled by $(+)$ and $(-)$,  respectively.

\begin{figure}
    \centering
    \includegraphics[width=.9\textwidth]{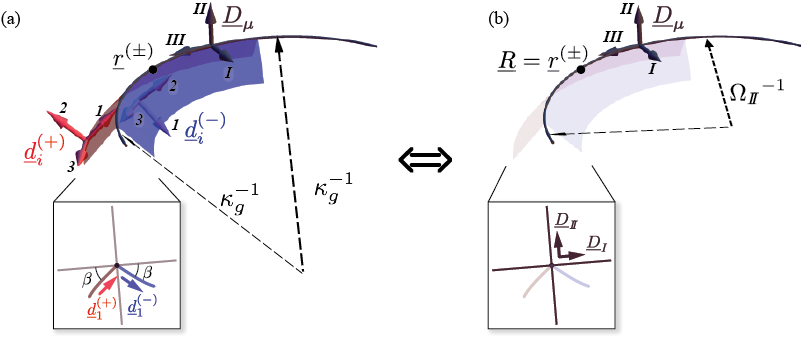}
    \caption{(a) A 3d configuration of the bistrip.  The flaps $(+)$
    and $(-)$ on both sides of the ridge are developable surface.
    Material frames $\underline{d}_{i}^{(\pm)}$ are attached to them.
    (b) The equivalent rod model makes use of the common ridge as the
    centerline, $\underline{R} = \underline{r}$, and of the bisecting
    frame $\underline{D}_{\mu}$ as the material frame; the ridge angle
    $\beta$ is viewed as an internal degree of freedom.  Note that the
    conserved geodesic curvature $\kappa_{\mathrm{g}}$ is measured
    along the tangent plane to the flaps, while the curvature
    strain $\Omega_{\II}$ of the centerline is measured in the plane
    spanned by $\underline{D}_{I}$ and $\underline{D}_{\III}$.
    By equation~(\ref{eq:kappaG-OmegaII-Recap}), $\Omega_{\II} \geq 
    \kappa_{\mathrm{g}}$, and the ratio of these curvatures sets the 
    ridge angle $\beta$.}
    \protect\label{fig:fold1}
\end{figure}

A typical 3d configuration of the bistrip is shown in
figure~\ref{fig:fold1}.
 To apply the analysis of
section~\ref{sec:singleStrip} to each of these flaps, we attach a
material frame $\underline{d}_{i}^{(\epsilon)}$, with $i=1,2,3$ and 
$\epsilon=\pm$ to them. Let us denote by $\underline{R}(s)$ the 
common ridge, and $s$ the arc-length along this ridge. We use this 
ridge as the centerline for both flaps: in the notations of the 
previous section, $\underline{r}^{(+)}(s) = \underline{r}^{(-)}(s) =
\underline{R}(s)$.

We observe that the tangent material vectors
$\underline{d}_{3}^{(\pm)}$ to both flaps are identical, since they
share the same centerline: by equation~(\ref{eq:tangent}),
$\underline{d}_{3}^{(+)}(s) = {\underline{r}^{(+)}}'(s) =
\underline{R}'(s) = {\underline{r}^{(-)}}'(s) =
\underline{d}_{3}^{(-)}(s)$.  This allows us to define the
\emph{bisecting frame} $\underline{D}_{\mu}(s)$, with $\mu =
I,\II,\III$ as follows.  Let us first define $\underline{D}_{\III}(s)$
to be the unit tangent to the ridge,
\begin{highlightequation}
    \begin{equation}
    \underline{R}'(s) = \underline{D}_{\III}(s)
    \textrm{,}
    \label{eq:midstripTangentCompatibility}
    \end{equation} 
\end{highlightequation}
which coincides with the other tangents, $\underline{D}_{\III}(s) =
\underline{d}_{3}^{(+)}(s) = \underline{d}_{3}^{(-)}(s)$.  The vector
$\underline{D}_{I}(s)$ is defined as the unit vector that bisects the
directions spanned by $\underline{d}_{1}^{(+)}(s)$ and
$\underline{d}_{1}^{(-)}(s)$, as shown in the insets of
figure~\ref{fig:fold1}.  Similarly, the vector
$\underline{D}_{\II}(s)$ is the unit vector that bisects the
directions spanned by $\underline{d}_{2}^{(+)}(s)$ and
$\underline{d}_{2}^{(-)}(s)$.  By construction, the bisecting frame is
an orthonormal frame,
\begin{equation}
    \underline{D}_{\mu}(s)\cdot \underline{D}_{\nu}(s) = \delta_{\mu\nu}
    \textrm{,}
    \label{eq:midstripOrthonormal}
\end{equation}
for any pair of indices, $\mu,\nu = I,\II,\III$.
As a result, the gradient of rotation of the bisecting frame
$\underline{D}_{\mu}$ can be measured by a Darboux vector
$\underline{\Omega}$ such that
\begin{highlightequation}
\begin{equation}
    \underline{D}_{\mu}'(s) = 
    \underline{\Omega}(s)\times \underline{D}_{\mu}(s)
    \textrm{.}
    \label{eq:darbouxForBisectingFrame}
\end{equation}
\end{highlightequation}

The star symbols in equation labels, as in
equations~(\ref{eq:midstripTangentCompatibility})
and~(\ref{eq:darbouxForBisectingFrame}), will be used to mark any
equation that defines the equivalent rod model.

Let us denote by $\beta$ half of the bending angle of the ridge, see
the inset of figure~\ref{fig:fold1}(a).  This angle can be defined as a
signed quantity if we use the orientation provided by the tangent
$\underline{D}_{\III} = \underline{R}'$: by convention $\beta$ is
positive in the figure.  In terms of the parameter $\beta$, the
dihedral angle at the ridge writes $\pi-2\,\beta$.

The local material frames $\underline{d}_{i}^{(\epsilon)}$ ($i=1,2,3$
and $\epsilon = \pm$) can be reconstructed in terms of the bisecting 
frame and of the angle $\beta$ as follows:
\begin{subequations}
    \label{eq:RotationBisectingFrame}
    \begin{align} 
	\underline{d}_{1}^{(\epsilon)}(s)&=\underline{D}_{I}(s)\,\cos\beta(s)
	+\epsilon\underline{D}_{\II}(s)\,\sin\beta(s)\\ 
	\underline{d}_{2}^{(\epsilon)}(s)&=-\epsilon\,\underline{D}_{I}(s)\,
	\sin\beta(s) +\underline{D}_{\II}(s)\, \cos\beta(s)\\
	\underline{d}_{3}^{(\epsilon)}(s)&=\underline{D}_{\III}(s).
    \end{align}
\end{subequations}
By this equation, the entire bistrip can be reconstructed in terms of
$\underline{R}$, $\underline{D}_{\mu}$ and $\beta$.  Therefore,
\emph{we use the ridge $\underline{R}$, the bisecting frame
$\underline{D}_{\mu}$ and the ridge angle $\beta$ as the main
unknowns}.  We shall show that the bistrip is equivalent to a rod
having a centerline $\underline{R}(s)$ and a material frame
$\underline{D}_{\mu}(s)$.  The kinematic
equations~(\ref{eq:midstripTangentCompatibility}) shows that this
equivalent rod is \emph{effectively} an inextensible, Navier-Bernoulli
rod --- we emphasize that the cross-sections are not assumed to be
rigid, however, as already discussed in section~\ref{sec:singleStrip}.
The rest of this section is concerned about deriving the constitutive
law for the equivalent rod that captures the elasticity of the flaps
and of the ridge.

\subsection{Reconstruction of local strains}

By differentiating equation~(\ref{eq:RotationBisectingFrame}) with
respect to arc-length $s$, and identifying the result with 
equation~(\ref{eq:darbouxVectorDefinition}) defining the local 
strains $\underline{\omega}^{(\epsilon)}$ in each flap, $\epsilon = 
\pm$, we obtain the following expression of the local strains:
\begin{equation}
    \underline{\omega}^{(\epsilon)}(s) = 
    \underline{\Omega}(s) + \epsilon\,\beta'(s)\,\underline{D}_{\III}(s)
    \mathrm{.}
    \label{eq:reconstructDarbouxVectors}
\end{equation}
Note that the second term is associated with the `internal' degree of
freedom $\beta$ associated with the ridge angle.

To use the constitutive laws for a single strip derived earlier, we
shall need the components of $\underline{\omega}^{(\epsilon)}$ in the
local material frame, which we denote by $\omega^{(\epsilon)}_{i} =
\underline{\omega}^{(\epsilon)}\cdot \underline{d}_{i}^{(\epsilon)}$
in the local frames.  Projecting the previous equation, we find
\begin{subequations}
    \label{eq:MaterialAndBisectingFrame}
   \begin{align}    
       \omega_{1}^{(\epsilon)}(s)&=\Omega_{I}(s)\,\cos\beta(s)+
       \epsilon\,\Omega_{\II}(s)\,\sin\beta(s)
       \label{eq:MaterialAndBisectingFrameRelation1}\\
       \omega_{2}^{(\epsilon)}(s)&=-\epsilon\,\Omega_{I}(s)\,\sin\beta(s)+
       \Omega_{\II}(s)\,\cos\beta(s)
        \label{eq:MaterialAndBisectingFrameRelation2}\\
       \omega_{3}^{(\epsilon)}(s)&=\Omega_{\III}(s)+\epsilon\,\beta'(s)
       \textrm{.}
        \label{eq:MaterialAndBisectingFrameRelation3}
    \end{align}
\end{subequations}
In the right-hand side, we denote by $\Omega_{\mu} =
\underline{\Omega}\cdot \underline{D}_{\mu}$ the projections of the
strain vector $\underline{\Omega}$ of the equivalent rod in its own
material (bisecting) frame $\underline{D}_{\mu}$, with $\mu =
I,\II,\III$.

\subsection{Ridge: internal stress, constitutive law}

Let us denote by $\underline{p}(s)$ and $\underline{q}(s)$ the force
and moment applied across the ridge, per unit arc-length
$\mathrm{d}s$, by the inner region $(-)$ onto the outer region $(+)$.
The outer flap feels a force $\underline{p}^{(+)}(s) =
+\underline{p}(s)$ and moment $\underline{q}^{(+)}(s) =
+\underline{q}(s)$.  By the principle of action-reaction, the inner
flap feels the opposite force and moment, $\underline{p}^{(-)}(s) =
-\underline{p}(s)$ and moment $\underline{q}^{(-)}(s) =
-\underline{q}(s)$.  We write this in compact form as
\begin{align}
    \underline{p}^{(\epsilon)}(s) = -\epsilon\, 
    \underline{p}(s),
    \label{eq:p-epsilon}
    \\
    \underline{q}^{(\epsilon)}(s) = -\epsilon\, 
    \underline{q}(s)
    \textrm{,}
    \label{eq:qepsilon-to-q}
\end{align}
for $\epsilon = \pm$.

We model the central fold as an elastic hinge.  The twisting moment
$\underline{q}(s)\cdot \underline{R}'(s)$ is therefore assumed to be a
function of the angle $\beta$:
\begin{equation}
    \underline{q}(s)\cdot \underline{R}'(s) = 
    Q_{\mathrm{r}}(2\,\beta)
    \emph{,}
    \label{eq:ridgeConstitutiveLaw}
\end{equation}
where $Q_{\mathrm{r}}$ is the constitutive law of the ridge.  By
convention, the argument of $Q_{\mathrm{r}}$ is the turning angle
$2\,\beta$ at the ridge, and not $\beta$.  This is motivated by the
fact that the work done by the ridge is
$(2\,\delta\beta)\,Q_{\mathrm{r}}(2\,\beta)$ when the parameter
$\beta$ is incremented by $\delta\beta$.

Assuming that the constitutive law of the ridge is linear, we write
\begin{equation}
    Q_{\mathrm{r}}(2\beta) = K_{\mathrm{r}}\times(2\,\beta - 
    2\,\beta_{\textrm{n}}),
    \label{eq:linearRidgeConstitutiveLaw}
\end{equation}
where $K_{\mathrm{r}}$ is the stiffness of the ridge, and
$\beta_{\textrm{n}}$ the natural value of the angle $\beta$.  By
creasing the strip, one induces irreversible deformations at the
ridge: this is modeled by changing the value of $\beta_{\mathrm{n}}$.

\subsection{Equations of equilibrium}

Equation~(\ref{eq:classicalKirchhoffEquilibrium-r}) expresses the
balance of force in each flap.  With our current notations, it can be
rewritten as ${\underline{n}^{(\epsilon)}}'(s) +
\underline{p}^{(\epsilon)}(s) = \underline{0}$, where $\epsilon = \pm$
labels the flaps and $\underline{n}^{(\epsilon)}$ denotes the internal
force in each flap. We define the total internal force 
$\underline{N}(s)$ in the bistrip,
\begin{equation}
    \underline{N}(s) = \underline{n}^{(+)}(s) + \underline{n}^{(-)}(s) 
    \textrm{.}
    \label{eq:totalInternalForce}
\end{equation}
By summing the local balance of forces and using the definition of 
$\underline{p}^{(\epsilon)}$ in equation~(\ref{eq:p-epsilon}), we 
find that the bistrip satisfies the global balance of forces
\begin{highlightequation}
\begin{equation}
    \underline{N}'(s) = \underline{0}
    \textrm{.}
    \label{eq:2Strips-BalanceOfForce}
\end{equation}
\end{highlightequation}
External force applied on the bistrip could be considered by adding a
term in the left-hand side, as in the classical theory of rods.

Let us now turn to the balance of moments, which can be written in
each flap as in equation~(\ref{eq:classicalKirchhoffEquilibrium-m}).
Recalling that the two flaps share the same tangent $\underline{r}' =
\underline{R}' = \underline{D}_{\III}$, we have
${\underline{m}^{(\epsilon)}}' + \underline{D}_{\III}\times
\underline{n}^{(\epsilon)} + \underline{q}^{(\epsilon)}
=\underline{0}$.  In terms of the total moment $\underline{M}(s)$ in
the bistrip, defined by
\begin{equation}
    \underline{M}(s) = \underline{m}^{(+)}(s) + \underline{m}^{(-)}(s)
    \textrm{,}
    \label{eq:totalInternalMoment}
\end{equation}
we write the global balance of moment in the bistrip as
\begin{highlightequation}
\begin{equation}
    \underline{M}'(s) + \underline{R}'(s)\times \underline{N}(s) = \underline{0}
    \textrm{.}
    \label{eq:2Strips-BalanceOfMoments}
\end{equation}
\end{highlightequation}

In equations~(\ref{eq:2Strips-BalanceOfForce})
and~(\ref{eq:2Strips-BalanceOfMoments}), we have recovered the
classical Kirchhoff equation for the equilibrium of thin rods: these
equations express the global balance of forces and moments in the
bistrip.

The global balance of moments~(\ref{eq:2Strips-BalanceOfMoments}) does
not involve the internal twisting moment due to the ridge,
$Q_{\mathrm{r}}$.  A second equation for the balance of moments can be
derived by projecting the local balance of moments in each strip onto
the shared tangent: ${\underline{m}^{(\epsilon)}}'\cdot \underline{R}'
+ \underline{q}^{(\epsilon)}\cdot \underline{R}' =0$.  Subtracting the
equations corresponding to $\epsilon = +$ and $\epsilon= -$, and
expressing $\underline{q}^{(\pm)}$ in terms of $Q_{\mathrm{r}}$ using
equations~(\ref{eq:qepsilon-to-q})
and~(\ref{eq:ridgeConstitutiveLaw}), we have
\begin{equation}
    \underline{\Delta}'(s)\cdot \underline{D}_{\III}(s)
    - Q_{\mathrm{r}}(2\,\beta(s))
    = 0
    \textrm{,}
    \label{eq:ridgeEquilibrium}
\end{equation}
where $\underline{\Delta}$ is half the difference of the internal moments:
\begin{equation}
    \underline{\Delta}(s) = \frac{\underline{m}^{(+)}(s)
    - \underline{m}^{(-)}(s)}{2}
    \textrm{.}
    \label{eq:differentialMoment}
\end{equation}
As suggested by the presence of the ridge moment $Q_{\mathrm{r}}$,
equation~(\ref{eq:ridgeEquilibrium}) expresses the balance of moments
at the ridge.  It can be viewed as the equation that sets the internal
degree of freedom $\beta$.

\subsection{Kinematic constraints applicable to the equivalent rod}

Two kinematic constraints are applicable in each flap $\epsilon =
\pm$: the geodesic constraint
$\mathcal{C}_{\mathrm{g}}^{(\epsilon)}=0$ and the developability
constraint $\mathcal{C}_{\mathrm{d}}^{(\epsilon)}=0$, see
equations~(\ref{eq:GaussCondition}) and~(\ref{eq:geodesicConstraint}).
Below, we express these constraints in terms of the centerline
$\underline{R}$, of the bisecting frame $\underline{D}_{\mu}$, and of
the ridge angle $\beta$.  This yields effective kinematic constraints
that are applicable to the equivalent rod.

Let us start by the geodesic constraint in
equation~(\ref{eq:geodesicConstraint}).  The geodesic curvature
$\kappa_{\mathrm{g}}$ has been interpreted in
figure~\ref{fig:FlatBistrip}, and is identical in both flaps:
$\omega_{2}^{(+)}(s) = \omega_{2}^{(-)}(s) = \kappa_{\mathrm{g}}$.  In
particular, the average of the local curvature reads
$\frac{1}{2}\,\left(\omega_{2}^{(-)}(s) + \omega_{2}^{(+)}(s)\right) =
\kappa_{\mathrm{g}}$.  Using
equation~(\ref{eq:MaterialAndBisectingFrameRelation2}), we can rewrite
the left-hand side in terms of the strain $\underline{\Omega}$ of the
equivalent rod:
\begin{highlightequation}
    \begin{equation}
    \Omega_{\II}(s)\,\cos\beta(s) = \kappa_{\mathrm{g}}
    \textrm{.}
    \label{eq:kappaG-OmegaII}
\end{equation}
By this kinematic constraint, the internal degree of freedom $\beta$
appears to be a function of the curvature strain $\Omega_{\II}$.  We
could eliminate $\beta$ in favor of $\Omega_{\II}$ using this
equation.  We shall instead view $\beta$ and $\Omega_{\II}$ as two
degrees of freedom subjected to the
constraint~(\ref{eq:kappaG-OmegaII}): this makes the final equations
easier to interpret.

A second constraint follows from the equality $\omega_{2}^{(+)}(s) =
\omega_{2}^{(-)}(s)$: when expressed in terms of $\underline{\Omega}$
as above, it reads $\Omega_{I}\,\sin\beta = 0$.  We shall ignore the
special case $\beta = 0$: as explained in
section~\ref{ssec:flat-is-pathological}, the bistrip is then on the boundary of
the space of configurations and the equations of equilibrium are
inapplicable anyway.  Under the assumption $\beta\neq 0$, we have:
\begin{equation}
	\Omega_I(s)=0
	\textrm{.}
	\label{eq:OmegaICancels}
\end{equation}
\end{highlightequation}

In view of the two constraints just derived, we can simplify the
expressions of the local strains given earlier in
equations~(\ref{eq:MaterialAndBisectingFrameRelation1})
and~(\ref{eq:MaterialAndBisectingFrameRelation3}):
\begin{subequations}
    \label{eq:MaterialAndBisectingFrame2}
   \begin{align}    
       \omega_{1}^{(\epsilon)}(s)&=\epsilon\,\Omega_{\II}(s)\,\sin\beta(s)
       \label{eq:MaterialAndBisectingFrameRelation21}\\
       \omega_{3}^{(\epsilon)}(s)&=\Omega_{\III}(s)+\epsilon\,\beta'(s)
        \label{eq:MaterialAndBisectingFrameRelation23}
	\textrm{.}
    \end{align}
\end{subequations}
The developability constraint in
equation~(\ref{eq:developabilityConstraint-KEliminiated}) can be 
simplified as well:
\begin{equation}
    \sigma^2(s)\,\lambda_{\mathrm{d}}^{(\epsilon)}(s)
	+ (\Omega_{\III}(s)+\epsilon\,\beta'(s))^2 = 0,
    \label{eq:devInTermsOfEpsilon}
\end{equation}
where we have introduced an auxiliary variable $\sigma(s) = \epsilon
\, \omega_{1}^{(\epsilon)}(s)$ which is given in terms of
$\underline{\Omega}$ by
\begin{highlightequation}
\begin{equation}
    \sigma(s)
    = \Omega_{\II}(s)\,
    \sin\beta(s)
    \textrm{.}
    \label{eq:sigma}
\end{equation}    
\end{highlightequation}

We note that the first term in equation~(\ref{eq:ridgeEquilibrium})
expressing the balance of moments at the ridge can be written in
coordinates as: $\underline{\Delta}'\cdot \underline{D}_{\III} =
(\underline{\Delta}\cdot \underline{D}_{\III})' - \Delta\cdot
\underline{D}_{\III}' = {\Delta_{\III}}' -
\underline{\Delta}\cdot(\underline{\Omega}\times
\underline{D}_{\III})$, where $\underline{\Omega}\times
\underline{D}_{\III} = \Omega_{\II}\,\underline{D}_{I}$ by the
constraint in equation~(\ref{eq:OmegaICancels}).  Here, we denote by
$\Delta_{\mu} = \underline{\Delta}\cdot \underline{D}_{\mu}$ and
$\Omega_{\mu} = \underline{\Omega}\cdot \underline{D}_{\mu}$ the
components of the differential internal moment $\underline{\Delta}$
and of the twist-curvature strain $\underline{\Omega}$ in the
bisecting frame.  We can therefore rewrite the equilibrium of the
ridge as
\begin{highlightequation}
\begin{equation}
    \Delta_{\III}'(s) - \Delta_{I}(s)\,\Omega_{\II} (s)
    -Q_{\mathrm{r}}(2\,\beta(s)) = 0\textrm{.} 
    \label{eq:ridgeEquilibriumInCoordinates}
\end{equation}
\end{highlightequation}

\subsection{Constitutive law}

To obtain a complete set of equations for the bistrip, we need the
expressions of the total internal moment $\underline{M}$ and of the
differential internal moment $\underline{\Delta}$ appearing in the
equations of equilibrium.  We derive the constitutive laws of the
bistrip below, by combining the local constitutive law in each flap,
and expressing them in terms of the twist-curvature strain
$\underline{\Omega}$ of the effective rod.

Let us denote the average over the two flaps $\epsilon = \pm$ by
angular brackets: $\langle f^{(\epsilon)} \rangle_{\epsilon} =
\frac{1}{2}(f_{-} + f_{+})$.  In terms of the Lagrange multipliers
$\lambda_{\mathrm{d}}^{(\epsilon)}$ associated with the developability
constraint in each flap, we define the following quantities:
\begin{subequations}
    \label{eq:AveragedModuli}
    \begin{align}
        b_{1}^+(s) & = \left\langle 1- 
	\left(\lambda_{\mathrm{d}}^{(\epsilon)}(s)\right)^2
	\right\rangle_{\epsilon}\\
        b_{1}^-(s) &  = \left\langle \epsilon\,\left(1- 
	\left(\lambda_{\mathrm{d}}^{(\epsilon)}(s)\right)^2\right)
	\right\rangle_{\epsilon}\\
        b_{3}^+(s) & = \left\langle 1- 
	\lambda_{\mathrm{d}}^{(\epsilon)}(s)
	\right\rangle_{\epsilon}\\
	b_{3}^-(s) & = \left\langle \epsilon\,\left(1- 
	\lambda_{\mathrm{d}}^{(\epsilon)}(s)\right)
	\right\rangle_{\epsilon}.
    \end{align}
\end{subequations}
Inserting the expression for $\lambda_{\mathrm{d}}^{(\epsilon)}$ found
in equation~(\ref{eq:devInTermsOfEpsilon}), we find explicit 
expressions for the auxiliary variables $b_{k}^\pm$:
\begin{subequations}
    \label{eq:explicitAverageModuli}
    \begin{highlightequation}
    \begin{align}
        b_{1}^+(\beta,\beta',\Omega_{\II},\Omega_{\III}) & = 
	1-\frac{1}{\sigma^{4}(\Omega_{\II},\beta)}\,({\Omega_{\III}}^4+6\,{\Omega_{\III}}^2\,{\beta'}^2+{\beta'}^4)\\
        b_{1}^-(\beta,\beta',\Omega_{\II},\Omega_{\III})  & =
	-\frac{4}{\sigma^{4}(\Omega_{\II},\beta)}\,({\Omega_{\III}}^3\,{\beta'}+{\Omega_{\III}}\,{\beta'}^3)\\
        b_{3}^+(\beta,\beta',\Omega_{\II},\Omega_{\III}) & =
	1+\frac{1}{\sigma^{2}(\Omega_{\II},\beta)}\,({\Omega_{\III}}^2+{\beta'}^2)\\
        b_{3}^-(\beta,\beta',\Omega_{\II},\Omega_{\III}) &  =
	\frac{2}{\sigma^{2}(\Omega_{\II},\beta)}\,{\Omega_{\III}}\,{\beta'}.
    \end{align}
    \end{highlightequation}
\end{subequations}

The Lagrange multipliers $\lambda_{\mathrm{g}}^{(\epsilon)}$
associated with the geodesic constraints are eliminated in favor of
their average $\Lambda_{+}(s) = \langle
\lambda_{\mathrm{g}}^{(\epsilon)}(s) \rangle_{\epsilon}$ and
half-difference $\Lambda_{-}(s) = \langle
\epsilon\,\lambda_{\mathrm{g}}^{(\epsilon)}(s) \rangle_{\epsilon}$:
for $\epsilon = \pm$, they can be reconstructed by
\begin{equation}
    \label{eq:LambdaGPlusMinus}
    \lambda_{\mathrm{g}}^{(\epsilon)}(s) = \Lambda_{+}(s)
    +\epsilon\,\Lambda_{-}(s)
    \textrm{.}
\end{equation}
We view $\Lambda_{+}(s)$ and $\Lambda_{-}(s)$ as quantities attached to
the equivalent rod: they are the Lagrange multipliers associated
to the two kinematic constraints~(\ref{eq:kappaG-OmegaII})
and~(\ref{eq:OmegaICancels}).

Let us now consider the local constitutive law in
equation~(\ref{eq:constitutiveReduced}), which provides the expression
of the internal moment $\underline{m}^{(\epsilon)}$ in each flap as a
function of $\lambda_{\mathrm{d}}^{(\epsilon)}$,
$\lambda_{\mathrm{g}}^{(\epsilon)}$
and $\omega_{2}^{(\epsilon)}$.
These quantities can be 
expressed in terms of the properties of the equivalent rod, using 
equations~(\ref{eq:AveragedModuli}),
(\ref{eq:LambdaGPlusMinus}) and
(\ref{eq:geodesicConstraint}), respectively.
This yields
\begin{equation}
    \underline{m}^{(\epsilon)} = 
    B\,\left(
    (b_{1}^+ +\epsilon \,b_{1}^-)
    \,\omega_{1}^{(\epsilon)}
    \,\underline{d}_{1}^{(\epsilon)}
    + 
    (\Lambda_{+} + \epsilon\,\Lambda_{-})
    \,\underline{d}_{2}^{(\epsilon)}
    +
    2\,(b_{3}^+ +\epsilon \,b_{3}^-)
    \,\omega_{3}^{(\epsilon)}
    \,\underline{d}_{3}^{(\epsilon)}
    \right).
    \nonumber
\end{equation}
Inserting the expressions of the local strains
$\omega_{1}^{(\epsilon)}$ and $\omega_{3}^{(\epsilon)}$ and of the
local frame $\underline{d}_{i}^{(\epsilon)}$ in
equations~(\ref{eq:MaterialAndBisectingFrame2})
and~(\ref{eq:RotationBisectingFrame}), we obtain the following
expressions of the internal moments $\underline{M}$ and
$\underline{\Delta}$ defined in
equations~(\ref{eq:totalInternalMoment})
as~(\ref{eq:differentialMoment}):
\begin{highlightequation}
\begin{equation}
    \left(
    \begin{array}{c}
	M_{I} \\ M_{\II} \\ M_{\III}  \\ \hline
	\Delta_{I} \\ 
	\Delta_{\III}
    \end{array}
    \right)
    =B\,\left(
    \begin{array}{cc|cc|c} 
	\sin (2\, \beta )\, b_1^- & 0 & 0 & -2\, \sin \beta & 0 \\
	2 \,\sin ^2\beta \, b_1^+ & 0 & 2 \,\cos \beta  & 0 & 0 \\
	0 & 4 \,b_3^+ & 0 & 0 & 4 \,b_3^- \\ \hline
	\cos \beta \, \sin \beta \, b_1^+ & 0 & -\sin \beta  & 0 & 0 \\
	0 & 2 \,b_3^- & 0 & 0 & 2 \,b_3^+																															   \end{array}
    \right)
    \cdot
    \left(
    \begin{array}{c}
	\Omega_{\II} \\ \Omega_{\III}   \\ \hline
	\Lambda_{+} \\ \Lambda_{-}   \\ \hline
	\beta'
    \end{array}
    \right)
    \label{eq:effectiveConstitutiveLaw}
\end{equation}
\end{highlightequation}

This equation is the main result of section~\ref{sec:bistrip}, and one
of the main results of our paper.  It yields the constitutive law of
the rod that is equivalent to the bistrip, and captures the details of
how the strip deforms at the `microscopic' scale $w$.  The
constitutive law is geometrically exact and handles large
deformations of the cross-section: the inextensibility of the strip is
treated exactly, the cross-sections of the flaps may bend
significantly, and the angle $\beta$ can change by a finite amount.
The constitutive law is non-linear because of the
geometry: the coefficients
$b_{i}^\pm(\beta,\beta',\Omega_{\II},\Omega_{\III})$ in the matrix
above depend non-linearly on the strains through
equations~(\ref{eq:explicitAverageModuli}).
By contrast, the underlying plate model makes use of a linear
constitutive law: the bending energy of the plate is quadratic with
respect to the strain, see equation~(\ref{eq:bendingEnergy2}).

This constitutive law~(\ref{eq:effectiveConstitutiveLaw}) depends on
the bending strain $\Omega_{\II}$ --- recall that the other bending
mode $\Omega_{I}$ is frozen by equation~(\ref{eq:OmegaICancels}) --- and on the
twisting strain $\Omega_{\III}$, like in the classical theory of thin
elastic rods.  It also depends on the internal degree of freedom
$\beta$ and on its derivative $\beta'$, and on the Lagrange
multipliers $\Lambda_{+}$ and $\Lambda_{-}$ associated with the 
two applicable constraints.

The expression of $\Delta_{\II}$ has been omitted in the constitutive
law~(\ref{eq:effectiveConstitutiveLaw}), as it does not appear in the
equations of equilibrium: it is absent from
equation~(\ref{eq:ridgeEquilibriumInCoordinates}) expressing the
balance of moments at the ridge.  For reference, its expression is $
    \Delta_{\II}(s) =     
    \sin ^2\beta(s) 
    \,b_1^-\big(\beta(s),\Omega_{\II}(s),\Omega_{\III}(s),\beta'(s)\big)\,\Omega_{\II}(s)
    +\cos \beta(s)\,\Lambda_{-}(s)$.

\subsection{Summary: effective rod model for a bistrip}
\label{sec:Summary}

We recapitulate the equations that govern the bistrip, collecting them
by family: these are all the equations that we have marked by a star
symbol so far.

\begin{itemize}
    
\item The main \emph{unknowns} of the model are (\emph{i}) the
centerline $\underline{R}(s)$ and a direct orthonormal frame
$\big(\underline{D}_{\mu}(s)\big)_{\mu=1,2,3}$, as usual in rod
theory; and (\emph{ii}) an `internal' degree of freedom, namely the
ridge angle $\beta(s)$.

\item The \emph{kinematic equation} defining the strain
$\underline{\Omega}$ reads
\begin{equation}
    \underline{D}_{\mu}'(s) = \underline{\Omega}(s)\times \underline{D}_{\mu}(s)
    \textrm{.}
    \label{eq:darbouxForBisectingFrame-Recap}
\end{equation}
This is the standard definition of the twist-curvature strain vector for
rods.

\item The following \emph{kinematic constraints} are applicable:
\begin{subequations}
    \begin{gather}
	\underline{R}'(s) = \underline{D}_{\III}(s)
	\textrm{,}
	\label{eq:midstripTangentCompatibility-Recap}\\ 
	\Omega_{\II}(s)\,\cos\beta(s) = \kappa_{\mathrm{g}}
	\textrm{,}
	\label{eq:kappaG-OmegaII-Recap}\\
	\Omega_I(s)=0
	\textrm{.}	
	\label{eq:OmegaICancels-Recap}
	\end{gather}
\end{subequations}    
Equation~(\ref{eq:midstripTangentCompatibility-Recap}) defines the
classical inextensible Euler-Bernoulli rod model: $s$ being a
Lagrangian variable, $|\underline{R}'(s)| = |D_{\III}(s)| = 1$ implies
inextensibility and $\underline{R}'\cdot \underline{D}_{\mu} =
\underline{D}_{\III}\cdot \underline{D}_{\mu} = 0$ for $\mu = I ,\II $
implies the absence of normal shear.  The two kinematic
constraints~(\ref{eq:kappaG-OmegaII-Recap}--\ref{eq:OmegaICancels-Recap})
are specific to the bistrip.

\item The \emph{equations of equilibrium} read
\begin{subequations}
	\begin{gather}
		\underline{N}'(s) = \underline{0}
    		\label{eq:2Strips-BalanceOfForce-Recap}\\
   	 	\underline{M}'(s) + \underline{R}'(s)\times 
		\underline{N}(s) = \underline{0}
    		\label{eq:2Strips-BalanceOfMoments-Recap}\\
		\Delta_{\III}'(s) - \Delta_{I}(s)\,\Omega_{\II} (s)-Q_{\mathrm{r}}(2\,\beta(s)) =0
           \label{eq:ridgeEquilibriumInCoordinates-Recap}
	\end{gather}
\end{subequations} 
Equations~(\ref{eq:2Strips-BalanceOfForce-Recap}--\ref{eq:2Strips-BalanceOfMoments-Recap})
are the classical Kirchhoff equations for rods.  The additional
equation~(\ref{eq:ridgeEquilibriumInCoordinates-Recap}) expresses the
balance of moments at the ridge, and sets the equilibrium value of the
internal degree of freedom $\beta$.

\item Finally, the constitutive law reads
\begin{equation}
    \left(
    \begin{array}{c}
	M_{I} \\ M_{\II} \\ M_{\III}  \\ \hline
	\Delta_{I} \\ 
	\Delta_{\III}
    \end{array}
    \right)
    =B\,\left(
    \begin{array}{cc|cc|c} 
	\sin (2\, \beta )\, b_1^- & 0 & 0 & -2\, \sin \beta & 0 \\
	2 \,\sin ^2\beta \, b_1^+ & 0 & 2 \,\cos \beta  & 0 & 0 \\
	0 & 4 \,b_3^+ & 0 & 0 & 4 \,b_3^- \\ \hline
	\cos \beta \, \sin \beta \, b_1^+ & 0 & -\sin \beta  & 0 & 0 \\
	0 & 2 \,b_3^- & 0 & 0 & 2 \,b_3^+																															   \end{array}
    \right)
    \cdot
    \left(
    \begin{array}{c}
	\Omega_{\II} \\ \Omega_{\III}   \\ \hline
	\Lambda_{+} \\ \Lambda_{-}   \\ \hline
	\beta'
    \end{array}
    \right)
    \textrm{,}
    \label{eq:effectiveConstitutiveLaw-Recap}
\end{equation}
where the secondary variables
$b_{k}^\pm(\beta,\beta',\Omega_{\II},\Omega_{\III})$, with $k=1,3$, 
were defined in equation~(\ref{eq:explicitAverageModuli}).

\end{itemize}

\subsection{Simplified constitutive law for nearly circular configurations}
\label{ssec:nearCircularConstitutiveLaw}

A much simpler version of the constitutive law can be derived, which
is applicable to near circular geometries.  We assume that the ridge
angle and the twist can be expanded as:
\begin{subequations}
    \label{eq:nearlyCircularExpansion}
    \begin{alignat}{2}
	\Omega_{\III}(s)&=\;\,0 &+ &\;\Omega_{\III}^1(s) + \cdots
	\label{eq:nearlyCircularExpansion-a}\\
	\beta(s)&=\;\beta_0&+&\;\beta_{1}(s) + \cdots
	\label{eq:nearlyCircularExpansion-b}
    \end{alignat}
\end{subequations}
Here, $\beta_{0}$ is a constant, to be specified later, and
$\beta_{1}(s)$ and $\Omega_{\III}^1(s)$ are assumed to be small.
Until the end of this section, we retain the linear terms
$\beta_{1}(s)$ and $\Omega_{\III}^1(s)$, but neglect higher-order
terms. 

Later on, we shall show that the expansions in
equation~(\ref{eq:nearlyCircularExpansion}) are applicable to the
analysis of circular configurations of the bistrip, and of their
stability.

When inserting the expansions in to
equation~(\ref{eq:explicitAverageModuli}), we find
\begin{equation}
    b_{1}^{+} = 1,\quad
    b_{3}^{+} = 1,\quad
    b_{1}^{-} = 0,\quad
    b_{3}^{-} = 0\textrm{,}
    \nonumber
\end{equation}
up to second-order terms: almost all the highly non-linear terms
disappear from the constitutive law --- a few of them are still
present as we do not assume the angle $\beta_{0}$ to be small.

To linear order, the constitutive law for $\Delta_{I}$ reads
\begin{equation}
    \Delta_{I}(s) = B\,\sin\beta(s)\,\left(
    \Omega_{\II}(s)\,\cos\beta(s) - \Lambda_{+}(s)
    \right)
    = B\,\sin\beta(s)\,\left(
    \kappa_{\mathrm{g}} - \Lambda_{+}(s)
    \right)
    \nonumber
\end{equation}
after using the geodesic constraint in
equation~(\ref{eq:kappaG-OmegaII-Recap}).  Inserting the constitutive
law $\Delta_{\III}(s) = 2\,B\,\beta_{1}'(s)$ into the balance of
moments at the ridge in
equation~(\ref{eq:ridgeEquilibriumInCoordinates-Recap}), inserting the
above expression of $\Delta_{I}(s)$ into the resulting expression, and
solving for $\Lambda_{+}(s)$, we find
\begin{equation}
     \Lambda_{+}(s) = \kappa_{\mathrm{g}} + \frac{
     \frac{1}{B}\,Q_{\mathrm{r}}(2\,\beta(s))-2\,\beta''(s)
     }{\Omega_{\II}(s)\,\sin\beta(s)}
     \textrm{.}
    \label{eq:linearizedCaseEliminateLambdaPlus}
\end{equation}

Inserting this expression into the constitutive
law~(\ref{eq:effectiveConstitutiveLaw-Recap}) for $\underline{M}$, and
retaining terms up to the linear order, we find
\begin{subequations}
    \label{eq:linearizedFinalConstitutive}
    \begin{align}
        M_{I}(s) & = \Lambda_{I}(s)
        \label{eq:linearizedFinalConstitutive-I}\\
	M_{\II}(s) & = M_{\II}^\mathrm{nc}(\Omega_{\II}(s),\beta(s),\beta''(s))
        \label{eq:linearizedFinalConstitutive-II}\\
        M_{\III}(s) &  =
	4\,B\,\Omega_{\III}(s)
	\textrm{.}
        \label{eq:linearizedFinalConstitutive-III} \\
	\intertext{Here, $M_{\II}^\mathrm{nc}$ denotes the constitutive law in 
	bending, in the nearly circular case:}
	M_{\II}^\mathrm{nc}(\Omega_{\II},\beta,\beta'') & = 2\,B\,
	\Omega_{\II} + \frac{2\,B}{\Omega_{\II}\,\tan\beta}\, \left(
	\frac{1}{B}\,Q_{\mathrm{r}}(2\,\beta) - 2\,\beta'' \right)
	\label{eq:linearizedFinalConstitutive-M2nc}
	\textrm{.}
    \end{align}
\end{subequations}

In equation~(\ref{eq:linearizedFinalConstitutive-I}), we have denoted
by $\Lambda_{I}(s)=-2\,B\,\sin\beta(s)\,\Lambda_{-}(s)$ the right-hand
side.  This $\Lambda_{I}(s)$ is viewed as the Lagrange multiplier
associated with the constraint $\Omega_I=0$, and is used in
replacement of the other unknown $\Lambda_{-}$.  Note that $\beta$ can
be viewed in equation~(\ref{eq:linearizedFinalConstitutive-II}) as a
function of $\Omega_{\II}$ through the
constraint~(\ref{eq:kappaG-OmegaII-Recap}): because of the presence
of the factor $\beta''$, $M_{\II}$ depends on the first and second
derivatives of $\Omega_{\II}$ as well, and the constitutive law is of
the second-gradient type.

The constitutive law~(\ref{eq:linearizedFinalConstitutive}) is
considerably simpler than the fully non-linear one in
equation~(\ref{eq:effectiveConstitutiveLaw-Recap}).  It incorporates
the equilibrium of the ridge: there is no need to use
equation~(\ref{eq:ridgeEquilibriumInCoordinates-Recap}) when we use
the above constitutive law.  This effective rod model resembles a
classical rod model: the twist mode is governed by a classical linear
constitutive law, $M_{\III} = 4\,B\,\Omega_{\III}$; the internal
degree of freedom $\beta$ can be viewed a function of the bending
strain $\Omega_{\II}$ by the
constraint~(\ref{eq:kappaG-OmegaII-Recap}); the constitutive law in
bending is non-linear and of second-gradient type, see
equation~(\ref{eq:linearizedFinalConstitutive-M2nc}); the curvature
$\Omega_{I} = 0$ is frozen and the corresponding bending moment $M_{I}
= \Lambda_{I}$ is a Lagrange multiplier.

As we shall show, this simple constitutive is applicable to the
analysis of the circular solutions in section~\ref{sec:circular}, and
to their stability in section~\ref{sec:linearStability}.  To compute
the post-buckled solution in section~\ref{sec:AUTO}, however, we shall
revert to the fully non-linear constitutive law of the previous 
section.

\section{Circular solutions}
\label{sec:circular}

The rest of the paper is concerned with the analysis of the equilibria
of a bistrip closed into a loop, see figure~\ref{fig:RodSection}.
This problem has been considered recently in~\cite{Dias2012a}: the
authors have shown that the planar configuration of the bistrip is
non-planar as it buckles into a 3d shape.  They observed that for very
small widths ($\kappa_{\mathrm{g}}w\ll1$) the dihedral angle is
unaffected by this buckling instability, and remains uniform and
constant in the post-buckled regime.  Revisiting this problem, we show
that (\emph{i}) it is a variant of the classical buckling analysis of an
elastic ring, (\emph{ii}) a simple expression for the buckling
threshold can be derived, (\emph{iii}) the conservation of the
dihedral angle in the post-buckled regime can be explained based on
symmetry considerations, and (\emph{iv}) the closed bistrip can
exhibit another, novel type of buckling instability.

\subsection{Preparation of the circular configuration}
\label{ssec:preparation}

We start by explaining in this section how the initial, circular state
of the bistrip is prepared.  The procedure is sketched in
figure~\ref{fig:residual}, and we invite the reader to perform the
following experiments himself or herself:
\begin{figure}[h]
    \centering
    \includegraphics[width=.65\textwidth]{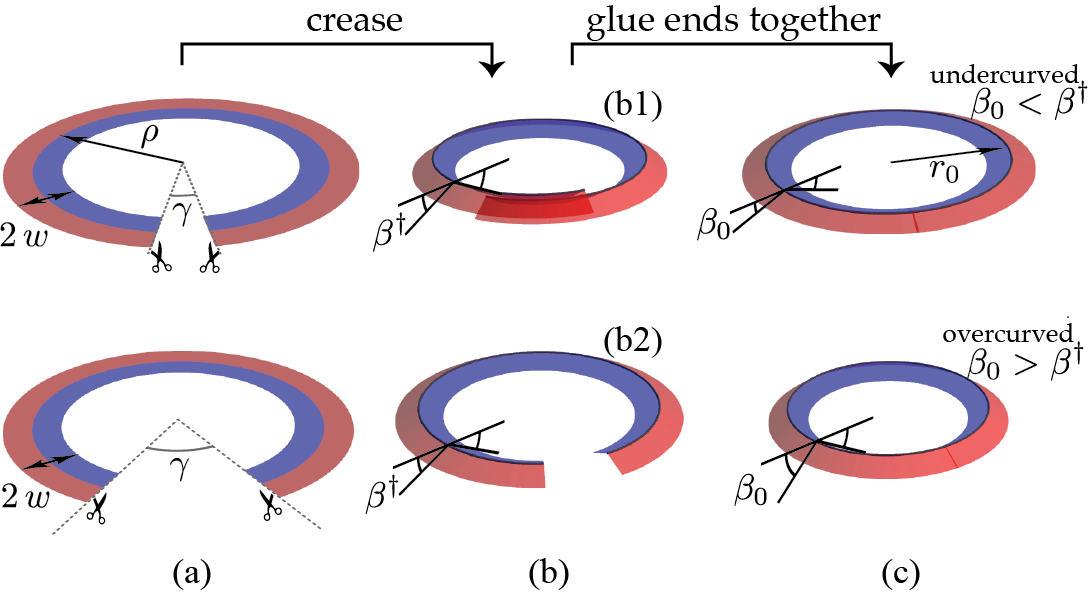}
    \caption{The circular bistrip is prepared by the following steps:
    (a) an annular region is cut out in a flat piece of paper, and a
    sector of angle $\gamma$ is removed from it.  (b) The ridge is
    creased: in the equations, this amounts to set the natural ridge
    angle $\beta_{\mathrm{n}}$ to a non-zero value.  As a result, the
    ridge angle $\beta$ has a non-zero equilibrium value $\beta^\dag$
    and the curvature of the centerline increases.  Depending on the
    value of $\gamma$, this can leads to an overlap (b1), or to a
    residual gap (b2).  (c) The bistrip is closed up by bringing the
    endpoints together.  Note that the circular configuration may be
    unstable, as studied later.}
    \label{fig:residual}
\end{figure}
\begin{itemize}
    \item An annular region of size $2\,w$ and mean radius $\rho$ is cut out in
a piece of paper. This sets the value of the geodesic curvature to
\begin{equation}
    \kappa_{\mathrm{g}} = \frac{1}{\rho}
    \textrm{.}
    \label{eq:kappag-from-rho}
\end{equation}
\item Next, a sector of angle $\gamma$ is removed, see
figure~\ref{fig:residual}(a).  The ridge is formed by creasing along the
central circle, as shown in figure~\ref{fig:residual}(b).  This amounts
to reset the value of the natural ridge angle $\beta_{\mathrm{n}}$
appearing in the constitutive
law~(\ref{eq:linearRidgeConstitutiveLaw}) to a non-zero value.  As a
result of this, the ridge angle takes on a value $\beta = \beta^\dag$
which is set by the the competition of the ridge energy (which is
minimum when $\beta_{\mathrm{n}}$) and of the bending energy of the
flaps (which is minimum when $\beta = 0$).  

\item Finally, the free
ends are glued together, as shown in figure~\ref{fig:residual}(c).  The
arc-length of the ridge is $L=(2\,\pi\,\rho - \gamma)\,\rho$, and so
the radius of curvature is
\begin{equation}
    r_{0} = \frac{L}{2\,\pi} = 
    \left(1-\frac{\gamma}{2\,\pi}\right)\,\rho
    \textrm{.}
    \label{eq:planar-r0}
\end{equation}
\end{itemize}

We analyze the circular configuration using the notations in
figure~\ref{fig:MaterialFrameFlat}:
\begin{figure}[tbp]
    \centerline{\includegraphics[width=.4\textwidth]{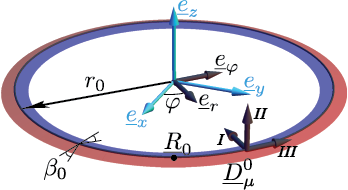}}
    \caption{Geometry of a circular configuration.}
    \protect\label{fig:MaterialFrameFlat}
\end{figure}
the $z$-axis is normal to the loop, and
the origin of coordinates is at the center of the circular ridge.  We
use the polar coordinates $(r,\varphi)$ in the plane $(x,y)$ containing
the ridge, and denote the polar basis by $\underline{e}_{r}(\varphi) =
\cos\varphi\,\underline{e}_{x} + \sin\varphi\,\underline{e}_{y}$ and
$\underline{e}_{\varphi}(\varphi) = -\sin\varphi\,\underline{e}_{x} +
\cos\varphi\,\underline{e}_{y}$.  Then, the centerline reads
\begin{equation}
    \underline{R}_{0}(s) = r_{0}\,\underline{e}_{r}\left(\frac{s}{r_{0}}
    \right)
    \textrm{,}
    \label{eq:planar-centerline}
\end{equation}
and the material frame
\begin{equation}
    \underline{D}_{I}^0(s) = 
    -\underline{e}_{r}\left(\frac{s}{r_{0}}\right),\quad
    \underline{D}_{\II}^0(s) = \underline{e_{z}},\quad
    \underline{D}_{\III}^0(s) = 
    \underline{e}_{\varphi}\left(\frac{s}{r_{0}}\right)
    \textrm{.}
    \label{eq:planar-material-frame}
\end{equation}
By the definition of the twist-curvature strain $\underline{\Omega}$ 
in equation~(\ref{eq:darbouxForBisectingFrame}), we have
\begin{equation}
    \underline{\Omega}_{0} = \Omega_{\II}^0\,\underline{e}_{z},\quad
    \textrm{where }\Omega_{\II}^0 = \frac{1}{r_{0}}
    \textrm{.}
    \label{eq:planar-darboux}
\end{equation}
This is compatible with the constraint $\Omega_{I}^0 = 
\underline{\Omega}^0\cdot \underline{D}_{I}^0 = 0$.  The twist is
also zero, $\Omega_{\III}^0 = 0$.  The ridge angle $\beta_{0}$ can be
found by equation~(\ref{eq:kappaG-OmegaII}) and it is uniform:
\begin{equation}
    \beta_{0} = 
    \cos^{-1}\left(\frac{\kappa_{\mathrm{g}}}{\Omega_{\II}^0}\right)
    = \cos^{-1}\left(1-\frac{\gamma}{2\,\pi}\right)
    \textrm{.}
    \label{eq:planar-beta0}
\end{equation}
Note that we can orient the $z$ axis so that it points in the same
direction as the ridge, as in figure~\ref{fig:residual} where both the
axis and the ridge point upwards.  Then, the angle is in the range
\begin{equation}
    0\leq \beta_{0}< \frac{\pi}{2}
    \textrm{.}
    \label{eq:beta0range}
\end{equation}

\subsection{Stress in the circular configuration}
\label{ssec:flat-is-pathological}

Since $\beta(s) = \beta_{0}$ is constant and $\Omega_{\III}(s) = 0$
cancels, the expansion postulated in
equation~(\ref{eq:nearlyCircularExpansion}) holds, with $\beta_{1} (s)
=0$ and $\Omega_{\III}^1(s) = 0$.  Therefore, the simplified
constitutive law in equation~(\ref{eq:linearizedFinalConstitutive}) is
applicable:
\begin{align}
    M_{I}^0(s) & = \Lambda_{I} \nonumber\\
    M_{\II}^0(s) & = M_{\II}^\mathrm{nc}\left(
    \frac{1}{r_{0}},\beta_{0},0\right)
    \nonumber\\
    M_{\III}^0(s) & = 0 
    \textrm{.}
    \nonumber
\end{align}
We seek a cylindrically symmetric solution, and the Lagrange
multiplier $\Lambda_{I}$ is therefore assumed to be
constant.

By the balance of forces in 
equation~(\ref{eq:2Strips-BalanceOfForce-Recap}), the internal force 
is constant, $\underline{N}_{0}(s) = \underline{N}_{0}$. By the 
balance of moments in 
equation~(\ref{eq:2Strips-BalanceOfMoments-Recap}), we find that 
$\Lambda_{I}$ and $\underline{N}_{0}$ both cancel. Therefore, the 
internal stress in the circular bistrip reads:
\begin{subequations}
    \label{eq:prestress}
    \begin{align}
        \underline{N}^0(s) & = \underline{0}
        \label{eq:forcePreStress} \\
        \underline{M}^0(s) & = M_{\II}^0\,\underline{e}_{z}
	\textrm{,}
        \label{eq:momentPreStress-TooGeneral}
    \end{align}
\end{subequations}
where the internal bending moment reads
\begin{equation}
    M_{\II}^0 =
    M_{\II}^\mathrm{nc}\left(
    \frac{1}{r_{0}},\beta_{0},0\right)=
    \frac{2\,B}{r_{0}}\,\left(
    1 +
    \frac{1}{\tan\beta_{0}}\,\frac{Q_{0}\,{r_{0}}^2}{B}
    \right)
    \textrm{.}
    \label{eq:momentPreStress-Explicit}
\end{equation}
Here, we have introduced the shorthand notation $Q_{0}$ for the ridge 
moment in the circular state:
\begin{equation}
    Q_{0} = Q_{\mathrm{r}}(2\,\beta_{0})
    \textrm{.}
    \label{eq:Q0notation}
\end{equation}
The first contributions to $M_{\II}^0$ in
equation~(\ref{eq:momentPreStress-Explicit}) is proportional to $B$
and comes from the elasticity of the flaps.  The second
term, proportional to $Q_{0}$, comes from the elasticity of the ridge.

The sign of the internal bending moment $M_{\II}^0$ is crucial for the
stability of the ring.  It can be positive or negative, depending on
how much the ridge has been creased (term $Q_{0}$) and how large the
ridge angle is (term $\beta_{0}$).  Following
reference~\cite{Mouthuy2012}, the circular solution is said to be
\emph{undercurved} when the internal moment $M_{\II}^0$ tends to
increase its curvature, and \emph{overcurved} in the opposite case:
\begin{equation}
    \begin{cases}
        M_{\II}^0<0 &\; : \; \textit{undercurved} \\
        M_{\II}^0>0 &\; : \; \textit{overcurved}
    \end{cases}
    \label{eq:under-vs-over-curved}
\end{equation}
The undercurved case corresponds to figure~\ref{fig:residual}(b1):
before they are glued together, the ends are overlapping; to close up
the ring, one has to decrease its curvature below its natural value,
making the ring wider and flatter: $\beta_{0}< \beta^\dag$.  The
overcurved case corresponds to the figure~\ref{fig:residual}(b2):
before the ends of the ring are glued together, they are separated by
a gap: closing up the ends involves decreasing the curvature below its
natural value, making the ring narrower and the ridge angle larger,
$\beta_{0}>\beta^\dag$.

We note that the stress in equation~(\ref{eq:prestress}) becomes
singular when $\beta_{0}=0$.  To avoid this difficulty, we shall
assume
\begin{equation}
    \beta_{0}\neq 0
    \textrm{.}
    \label{eq:beta0-cannot-cancel}
\end{equation}
The case $\beta_{0} = 0$ is pathological because the circular solution
then sits on the \emph{boundary} of the space of admissible
configuration, not in the interior.  Indeed, the ridge curvature
$\Omega_{\II}$ is at its maximum value, $\kappa_{\mathrm{g}}$: this
curvature cannot vary to first order, and as a result of this the
in-plane projection of the ridge deforms rigidly.
The equations of equilibrium that we derived are not applicable in
this special case.  To study the case $\beta_{0} = 0$, we would need
to relax constraints and consider an extensible plate model.

\section{Linear stability of the circular solutions}
\label{sec:linearStability}

In reference~\cite{Dias2012a}, non-planar configurations of the closed
bistrip were observed using a paper model; post-buckled solutions were
also calculated, showing striking similarities with the experimental
patterns.  A typical picture of a paper model is reproduced in our
figure~\ref{fig:RodSection}.

Here, we show that these shapes are produced by a buckling instability
affecting the circular solutions of section~\ref{sec:circular}.  Our
model allows us to identify the stress that causes this instability:
this is simply the prestress in equation~(\ref{eq:prestress}).  The
eigenmodes and the buckling threshold are calculated analytically.
The selection of the azimuthal wavenumber (number of bumps) is
explained.  A second family mode of buckling is pointed out, and
demonstrated experimentally.

\subsection{Parameterization of the buckling modes}

The buckling modes are parameterized by three functions, $\Big\{
\hat{\psi}_{I}(s), \hat{\psi}_{\II}(s), \hat{\psi}_{\III}(s) \Big\}$,
which are the components in the undeformed basis of the infinitesimal
rotation vector,
\begin{equation}
    \underline{\hat{\psi}}(s) = \sum_{\mu=I}^{\III }
    \hat{\psi}_{\mu}(s)\,\underline{D}_{\mu}^0(s)
    \textrm{,}
    \label{eq:reconstructLocalRotation}
\end{equation}
where hats denote perturbations, \emph{i.e.}\ small increments.  The
infinitesimal rotation $\underline{\hat{\psi}}(s)$ is used to
reconstruct the perturbed material frame by
\begin{equation}
    \underline{D}_{\mu}(s) = \underline{D}_{\mu}^0(s) + 
    \underline{\hat{\psi}}(s)\times \underline{D}_{\mu}^0(s)
    \textrm{.}
    \label{eq:stab-infinitesimalRotation}
\end{equation}
The perturbed centerline can then be found by integration of the
equation $\underline{R}' = \underline{D}_{\III}$, up to a constant of
integration which is an important rigid-body translation.

\subsection{Infinitesimal perturbation to the twist and curvature}

To compute the perturbed strain vector $\underline{\Omega}$, we take
the derivative of equation~(\ref{eq:stab-infinitesimalRotation}) and
use equation~(\ref{eq:darbouxForBisectingFrame}): 
$\underline{\Omega}(s)\times \underline{D}_{\mu}(s) = 
\underline{\Omega}_{0}(s) \times 
(\underline{D}_{\mu}(s)-\underline{\hat{\psi}}(s) \times 
\underline{D}_{\mu}^0(s)) + \underline{\hat{\psi}}'(s) \times 
\underline{D}_{\mu}^0(s) + \underline{\hat{\psi}}(s) \times 
(\underline{\Omega}_{0}(s)\times \underline{D}_{\mu}^0(s))$. 
Rearranging the terms and using Jacobi's identity, we derive the 
following 
expression for the perturbation $\underline{\hat{\Omega}}(s) = 
\underline{\Omega}(s) - \underline{\Omega}_{0}(s)$ to the strain 
vector:
\begin{equation}
    \underline{\hat{\Omega}}(s) = \underline{\hat{\psi}}'(s) - 
    \underline{\Omega}_{0}(s) \times \underline{\hat{\psi}}(s)
    \textrm{.}
    \label{eq:rotationsCompatibility}
\end{equation}
This equation expresses the geometric compatibility of the increment
of rotation $\underline{\hat{\psi}}(s)$ and of strain
$\underline{\hat{\Omega}}(s)$, and is well-known,
see~\cite{Chouaieb-Kirchhoffs-problem-of-helical-solutions-of-uniform-2003}
and~\cite[eq.~3.51]{Audoly-Pomeau-Elasticity-and-geometry:-from-2010} 
for instance. Its right-hand side can be interpreted as the co-moving derivative of
$\underline{\hat{\psi}}$ in the undeformed material frame
$\underline{D}_{\mu}^0$, which shows that
\begin{equation}
    \underline{\hat{\Omega}}(s) = \sum_{\mu = I}^{\III}
    \hat{\psi}_{\mu}'(s)\,\underline{D}_{\mu}^0(s)
    \textrm{.}
    \label{eq:hatOmegaVectorInComponents}
\end{equation}

We denote by $\hat{\Omega}_{\mu}(s)$ the first-order perturbation to
the strain caused by the perturbation:
\begin{equation}
    \hat{\Omega}_{\mu}(s) = 
    \Omega_{\mu}(s) - \Omega_{\mu}^0 (s) =
    \underline{\Omega}(s)\cdot \underline{D}_{\mu}(s)
    -
    \underline{\Omega}_{0}(s)\cdot \underline{D}_{\mu}^0(s)
    \textrm{.}
    \label{eq:defineOmegaHatMu}
\end{equation}
Neglecting second-order terms, we can write this variation of a
product as $\hat{\Omega}_{\mu} = \underline{\hat{\Omega}}\cdot
\underline{D}_{\mu}^0 + \underline{\Omega}_{0}\cdot
(\underline{\hat{\psi}}\times \underline{D}_{\mu}^0)$.  The first is 
given by
equation~(\ref{eq:hatOmegaVectorInComponents}), and we have
\begin{equation}
    \hat{\Omega}_{\mu}(s) = \hat{\psi}_{\mu}'(s)
    +
    \underline{\Omega}_{0}(s)\cdot
    (\underline{\hat{\psi}}(s)\times \underline{D}_{\mu}^0(s))
    \textrm{.}
\end{equation}
Inserting
the expression of $\underline{\Omega}_{0}$ in
equation~(\ref{eq:planar-darboux}), this yields
\begin{subequations}
    \label{eq:strainIncrements}
    \begin{align}
        \hat{\Omega}_{I}(s) &= \hat{\psi}_{I}'(s) +
	\frac{1}{r_{0}}\,\hat{\psi}_{\III}(s)
        \label{eq:strainIncrements-I}\\
        \hat{\Omega}_{\II}(s) &= \hat{\psi}_{\II}'(s)
        \label{eq:strainIncrements-II}\\
	\hat{\Omega}_{\III}(s) &= \hat{\psi}_{\III}'(s) -
	\frac{1}{r_{0}}\,\hat{\psi}_{I}(s)
        \label{eq:strainIncrements-III}
    \end{align}
\end{subequations}

The curvature $\Omega_{I}(s)$ being frozen by
equation~(\ref{eq:OmegaICancels-Recap}), its perturbation is zero,
$\hat{\Omega}_{I}(s) = 0$.  The rotation $\hat{\psi}_{\III}$ can then
be eliminated from equation~(\ref{eq:strainIncrements-I}):
\begin{equation}
    \hat{\psi}_{\III}(s) = - r_{0}\,\hat{\psi}_{I}'(s)
    \textrm{.}
    \label{eq:eliminatePsi3}
\end{equation}
Inserting into equation~(\ref{eq:strainIncrements-III}), we find
\begin{equation}
    \hat{\Omega}_{\III}(s) = 
    -\frac{1}{r_{0}}\,\left({r_{0}}^2\,\hat{\psi}_{I}''(s) +
	\hat{\psi}_{I}(s)\right)
	\textrm{.}
    \label{eq:strainIncrements-III-v2}
\end{equation}

\subsection{Azimuthal wavenumber}

Given the cylindrical invariance of the base solution, we seek
buckling modes that depend harmonically on the polar variable $\varphi
= \frac{s}{r_{0}} = \Omega_{\II}^0\,s$.  These buckling modes are
indexed by an integer wavenumber $n\geq 0$,
\begin{equation}
    \left(
   \hat{\psi}_{I}(s),
    \hat{\psi}_{\II}(s),
    \hat{\psi}_{\III}(s)
    \right) = 
    \left(
    \hat{\Psi}_{I},
    \hat{\Psi}_{\II},
    \hat{\Psi}_{\III}
    \right)\,e^{\frac{i\,n\,s}{r_{0}}}
    \textrm{.}
    \label{eq:stability-eigenmodes}
\end{equation}
The coefficients in the parenthesis in the right-hand side are the
complex amplitudes of the infinitesimal rotation.

By a classical argument, the cases $n=0$ and $n=1$, which correspond
to rigid-body rotations, are ruled out.  Indeed, when $n=0$,
$\hat{\Omega}_{I} = 0$ by the constraint, $\hat{\Omega}_{\II} = 0$ by
equation~(\ref{eq:strainIncrements-II}), and $\hat{\Omega}_{\III}$ is
proportional to $\hat{\psi}_{I}$ by
equation~(\ref{eq:strainIncrements-III-v2}); the condition that the
centerline closes up after one turn requires that the constant value
of $\hat{\psi}_{I}$ is zero.  As a result, all the strain components
$\Omega_{\mu}$ stay unperturbed when $n=0$, which corresponds to a
rigid-body motion of the bistrip.  A similar argument shows that $n=1$
corresponds to a rigid-body motion of the bistrip as well.  Therefore,
we only consider azimuthal wavenumbers $n$ in the buckling analysis
such that
\begin{equation}
    n\geq 2
    \textrm{.}
    \label{eq:n-larger-than-2}
\end{equation}

\subsection{Linearized equilibrium}

Linearizing the balance of
forces~(\ref{eq:2Strips-BalanceOfForce-Recap}), we have
$\underline{\hat{N}}'(s) = \underline{0}$, and so the perturbation to
the internal force is a constant vector, $\underline{\hat{N}}(s) =
\underline{0}$.  We know $\underline{N}_{0}(s) = \underline{0}$ from
equation~(\ref{eq:forcePreStress}), and therefore the total internal
force reads $\underline{N}(s) = \underline{\hat{N}}$ to first order in
the perturbation.  Inserting into the balance of
moments~(\ref{eq:ridgeEquilibriumInCoordinates-Recap}) and retaining
first order terms, we find
\begin{equation}
    \underline{\hat{M}}'(s) = -
    \underline{D}_{\III}^0(s)\times \underline{\hat{N}}
    \textrm{.}
    \label{eq:linearizedBalanceOfMoments}
\end{equation}
Since we study buckling modes that are pure Fourier modes, the
components of the vector $\underline{\hat{M}}'(s)$ in the unperturbed
frame $\underline{D}_{\mu}^0$ all depend on $s$ as
$\exp\left(\frac{i\,n\,s}{r_{0}}\right)$ with $n\geq 2$.  By contrast,
the components of the right-hand side $( -
\underline{D}_{\III}^0(s)\times \underline{\hat{N}})$ in the
unperturbed frame $\underline{D}_{\mu}^0$, which can be computed explicitly,
have only two non-zero Fourier components, with wavevectors $0$ and $1$.
Therefore, we conclude that both sides of
equation~(\ref{eq:linearizedBalanceOfMoments}) must cancel: 
$\underline{\hat{M}}'(s) = \underline{0}$. Integrating, we find that 
$\underline{\hat{M}}(s)$ is a constant. For the components of 
$\underline{\hat{M}}(s)$ in the material frame to be harmonic with 
$n\geq 2$, this constant must in fact be zero:
\begin{equation}
    \underline{\hat{M}}(s) = \underline{0}
    \textrm{.}
    \label{eq:MConstantByBuckling}
\end{equation}
We have just shown that the buckling modes leave the internal force 
and moment constant, to first order.

\subsection{Linearized constitutive law}

In equation~(\ref{eq:defineOmegaHatMu}), we have assumed that the
bending and twist strain can be expanded as $\Omega_{\mu}(s) =
\Omega_{\mu}^0(s) + \hat{\Omega}_{\mu}(s)$, with $\mu = \II,\III$.
Using the unperturbed strain $\Omega_{\mu}^0(s)$ given earlier in 
equation~(\ref{eq:planar-darboux}), this yields
\begin{subequations}
    \begin{alignat}{2}
	\Omega_{\II}(s)&=\Omega_{\II}^0\;&+&\;\hat{\Omega}_{\II}(s)\\
	\Omega_{\III}(s)&=\;\,0 &+ &\;\hat{\Omega}_{\III}(s)
	\label{eq:expansionAgain-a}
	\textrm{.}
    \end{alignat}
Using the geodesic constraint in
equation~(\ref{eq:kappaG-OmegaII-Recap}), we can derive the expansion
for the ridge angle $\beta$, from that of $\Omega_{\II}$:
\begin{equation}
    \beta(s) = \beta_{0} \;+\;\hat{\beta}(s) 
    \label{eq:expansionAgain-b}
\end{equation}
where
\begin{equation}
    \hat{\beta}(s) = 
    \frac{r_{0}}{\tan\beta_{0}}\,\hat{\Omega}_{\II}(s)
    \textrm{.}
    \label{eq:geodesicConstraintsLinearized}
\end{equation}
\end{subequations}

Comparison of
equations~(\ref{eq:expansionAgain-a}--\ref{eq:expansionAgain-b}) with
equations~(\ref{eq:nearlyCircularExpansion-a}--\ref{eq:nearlyCircularExpansion-b})
shows that the linear stability analysis involves exactly the type of
expansion that was postulated in
section~\ref{ssec:nearCircularConstitutiveLaw}, which we dubbed the
`nearly circular' case.  Therefore, \emph{we can use the simplified
constitutive law}~(\ref{eq:linearizedFinalConstitutive}) in the linear
stability analysis. Linearizing this constitutive law, we get
\begin{subequations}
    \label{eq:linearizedConstitutiveLaw}
    \begin{align}
        \hat{M}_{\II} & = 
	\frac{\partial 
	M_{\II}^\mathrm{nc}(\Omega_{\II}^0,\beta_{0},0)}{\partial 
	\Omega_{\II}}\,\hat{\Omega}_{\II}
	+
	\frac{\partial 
	M_{\II}^\mathrm{nc}(\Omega_{\II}^0,\beta_{0},0)}{\partial 
	\beta}\,\hat{\beta} 
	+
	\frac{\partial 
	M_{\II}^\mathrm{nc}(\Omega_{\II}^0,\beta_{0},0)}{\partial 
	\beta''}\,\hat{\beta}'' 
	\label{eq:linearizedConstitutiveLaw-2}\\
        \hat{M}_{\III} & = 4\,B\,\hat{\Omega}_{\III}
        \label{eq:linearizedConstitutiveLaw-3}
    \end{align}
\end{subequations}
Note that we have not included the linearized constitutive for
$\hat{M}_{I}$, as we are not interested in reconstructing the Lagrange
multiplier $\hat{\Lambda}_{I}$.  The quantities in the left-hand side
are the perturbations to the strain components $\hat{M}_{\mu} =
\underline{M}\cdot\underline{D}_{\mu} -
\underline{M}_{0}\cdot\underline{D}_{\mu}^0$. The perturbation to the 
internal moment $\underline{\hat{M}}$ can be written in terms of them 
as
\begin{multline}
    \underline{\hat{M}}(s) = \underline{M}(s) - \underline{M}_{0}(s)
    = 
    \sum_{\mu=I}^{\III}(\hat{M}_{\mu}(s)+M_{\mu}^0)\,\underline{D}_{\mu}(s)
    -M_{\mu}^0\,\underline{D}_{\mu}^{0}(s)
    \\
    =
    \sum_{\mu=I}^{\III}\hat{M}_{\mu}(s)\,\underline{D}_{\mu}(s)
    +
    \sum_{\mu=I}^{\III}M_{\mu}^0\,\underline{\psi}(s)\times 
    \underline{D}_{\mu}^0(s)
    = \sum_{\mu=I}^{\III}\hat{M}_{\mu}(s)\,\underline{D}_{\mu}^0(s)
    + \underline{\hat{\psi}}(s)\times \underline{M}^0
    \textrm{,}
\end{multline}
after dropping second-order terms.  By the balance of moments in
equation~(\ref{eq:MConstantByBuckling}), the left-hand side is zero.
Inserting the expression of $\underline{M}_{0}$ in
equation~(\ref{eq:momentPreStress-TooGeneral}) in the right-hand side,
and projecting onto the directions $\mu=\II$ and $\mu=\III$, we find
\begin{subequations}
     \label{eq:MmuChangeOfBasis}
    \begin{align}
        \hat{M}_{\II}(s) & = 0
         \label{eq:MmuChangeOfBasis-2}
	\\
        \hat{M}_{\III}(s) + M_{\II}^0\, \hat{\psi}_{I}(s) & = 0
	\textrm{.}
        \label{eq:MmuChangeOfBasis-3}
    \end{align}
\end{subequations}

\subsection{Centerline mode}

Eliminating $M_{\III}$ from
equations~(\ref{eq:linearizedConstitutiveLaw-3})
and~(\ref{eq:MmuChangeOfBasis-3}) and inserting the expression of
$\underline{\hat{\Omega}}_{\III}$ obtained by the kinematic
equation~(\ref{eq:strainIncrements-III-v2}), we find an eigenvalue
problem for the periodic function $\hat{\psi}_{I}$:
\begin{equation}
    -\frac{M_{\II}^0\,\hat{\psi}_{I}(s)}{4\,B}
    =
    -\frac{1}{r_{0}}\,\left(
    {r_{0}}^2\,\hat{\psi}_{I}''(s)+\hat{\psi}_{I}(s)
    \right)
    \textrm{.}
    \label{eq:eigenvalueProblemCenterline}
\end{equation}
Inserting the harmonic dependence on the arc-length $s$ given in
equation~(\ref{eq:stability-eigenmodes}), this leads to
\begin{equation}
    \left(\frac{M_{\II}^0\,r_{0}}{4\,B} 
    +(n^2-1)\right)\,\hat{\Psi}_{I} = 0
    \label{eq:centerlineModeCriterion-Pre}
    \textrm{.}
\end{equation}
When the factor in parentheses cancels, non-zero values of the 
rotation $\hat{\Psi}_{I}$ are possible. This corresponds to a family 
of buckling modes which we call the \emph{centerline mode}. 

In equation~(\ref{eq:centerlineModeCriterion-Pre}), equating the
factor in parenthesis to zero yields the critical value of the bending
prestress $M_{\II}^0$ where the ridge mode occurs.  The prestress
$M_{\II}^0$ is always negative when the parenthesis cancels in
equation~(\ref{eq:centerlineModeCriterion-Pre}), since $n\geq 2$.
Therefore, \emph{the centerline buckling is only possible in the
undercurved case.}

This prestress is
itself a function of the natural ridge angle $\beta_{\mathrm{n}}$ by
equation~(\ref{eq:momentPreStress-Explicit}).  Inserting this
function, we find an equation for the critical value of the natural
ridge angle $\beta_{\mathrm{n}} =
\beta_{\mathrm{n},\mathrm{crit}}^\mathrm{ctl}(
\overline{K}_{\mathrm{r}},
\beta_{0},n)$ where the
centerline buckling mode occurs:
\begin{equation}
    \overline{K}_{\mathrm{r}}\,
    \frac{2\,\beta_{0}
    -
    2\,\beta_{\mathrm{n},\mathrm{crit}}^\mathrm{ctl}(\overline{K}_{\mathrm{r}},
    \beta_{0},n)}{\tan\beta_{0}}
    = 1 - 2\,n^2,\qquad n\geq 2 \textrm{.}
    \label{eq:centerlineModeCriterion}
\end{equation}
This
$\beta_{\mathrm{n},\mathrm{crit}}^\mathrm{ctl}(\overline{K}_{\mathrm{r}},
\beta_{0},n)$, is a function of the initial ridge angle $\beta_{0}$,
of the wavenumber
$n$, and of the dimensionless ridge stiffness
\begin{equation}
    \overline{K}_{\mathrm{r}} =\frac{{r_{0}}^2\,K_{\mathrm{r}}}{B}
    \textrm{.}
    \label{eq:dimensionlessRidgeStiffness}
\end{equation}

\begin{figure}[h]
     \centering
      \includegraphics[width=0.9\textwidth]{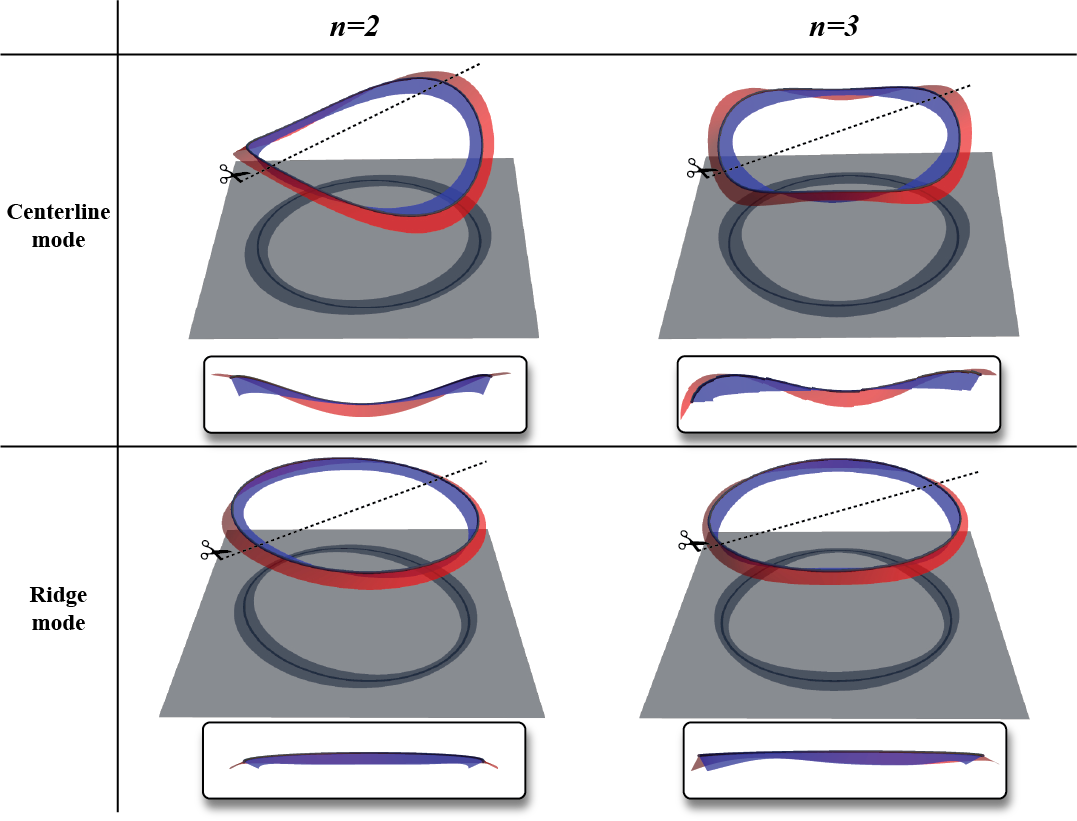}
     \caption{The two families of buckling modes for an initially
     circular configuration: centerline and ridge modes.  In each
     family, the first two modes are shown, corresponding to the
     wavenumbers $n=2$ and $3$.  In the shadows obtained by projection
     onto a normal plane, the centerline appears to be oscillating
     from one edge to the other in the centerline mode, but appears
     centered in the ridge mode.  Cuts along the dashed line are shown
     in the framed insets.  In the centerline mode, which occurs in pure twist,
     the central ridge goes out of plane, the dihedral angle is
     conserved and the cross-sections swing back and forth about the
     centerline.  By contrast, the ridge mode involves a modulation of
     the dihedral angle, and the central ridge stays planar.}
     \label{fig:CenterlineModes}
     \label{fig:RidgeModes}
 \end{figure}

The mode can be reconstructed by picking an infinitesimal value of
$\hat{\Psi}_{I}$ in equation~(\ref{eq:stability-eigenmodes}), and by
computing $\underline{\hat{\psi}}$ by
equation~(\ref{eq:eliminatePsi3}).  The twisting strain is then given
by equation~(\ref{eq:strainIncrements-III-v2}).  On the other hand, the
bending strain $\Omega_{\II}$, which is proportional to
$\hat{\psi}_{\II}$ remains zero: \emph{the centerline mode involves
buckling in pure twist}.  By equation~(\ref{eq:expansionAgain-b}), the
ridge angle remains unperturbed as well --- this is why we call it a
centerline mode.  The mode can be visualized by reconstructing the
perturbed material frame by
equation~(\ref{eq:stab-infinitesimalRotation}), and by integrating
along the tangent to find the deformed centerline.  In
figure~\ref{fig:CenterlineModes} the first two centerline modes, $n=2$
and $3$, are visualized.

In reference~\cite{Dias2012a}, post-buckled configurations of the
centerline mode have been observed both in experiments using paper
model and in simulations.  Here, we have shown that this mode occurs
by an instability very similar to classical instabilities for elastic
rings~\cite{Zajac-Stability-of-two-planar-loop-1962,%
Goriely-Twisted-Elastic-Rings-2006,%
Moulton2013}, and have calculated the buckling
threshold analytically.  Our equation predicts that $n=2$ is the first
unstable mode, when the prestress $M_{\II}^0$ is made more and more
negative.

\subsection{Analysis of the ridge mode}

We now proceed to analyze the second family of buckling modes, by
using the remaining equations derived in the beginning of this
section.  Combining the balance of
moments~(\ref{eq:MmuChangeOfBasis-2}) and the linearized constitutive
law~(\ref{eq:linearizedConstitutiveLaw-2}), we find a relation between
$\hat{\Omega}_{\II}$, $\hat{\beta}$ and $\hat{\beta}''$:
\begin{equation}
\label{eq:Change-Angle}
\hat{\beta}''=
\frac{K_\mathrm{r}-Q_0\csc(2\beta_0)}{B}\hat{\beta}+
\frac{1}{2\,r_0}\,\left(\tan\beta_0-\frac{r_0^2\,Q_0}{B}\right)
\,\hat{\Omega}_{\II}.
\end{equation}  
Then, we eliminate $\hat{\beta}$ using the linearized geodesic
constraint in equation~(\ref{eq:geodesicConstraintsLinearized}) and
the kinematic equation~(\ref{eq:strainIncrements-II}), and obtain an 
eigenvalue problem for the periodic function 
$\hat{\psi}_{\II}$:
\begin{equation}
\label{eq:strainIncrements-II-DiffEq}
\hat{\psi}_{\II}'''(s)=
\frac{1}{{r_{0}}^2}\,\left[
\frac{\tan^2(\beta_0)}{2}
+
\overline{K}_{\mathrm{r}}
-\frac{1}{2}\,
\left(
\tan\beta_{0}+\frac{1}{\sin\beta_{0}\,\cos\beta_{0}}\right)\,(2\,\beta_{0} - 
2\,\beta_{\mathrm{n}})
\,\overline{K}_{\mathrm{r}}
\right]
\,\hat{\psi}_{\II}'(s).
\end{equation}  
The harmonic dependence on the arc-length given in
equation~(\ref{eq:stability-eigenmodes}) is used again to solve
equation~(\ref{eq:strainIncrements-II-DiffEq}).  As earlier, this
yields the equation for the critical value of the natural angle
$\beta_{\mathrm{n}}$ at the ridge, which we denote by
$\beta_{\mathrm{n},\mathrm{crit}}^\mathrm{ridge}(\overline{K}_{\mathrm{r}},\beta_{0},n)$:
\begin{equation}
    \overline{K}_{\mathrm{r}}\,
    \frac{2\,\beta_{0}
    -
    2\,\beta_{\mathrm{n},\mathrm{crit}}^\mathrm{ridge}(\overline{K}_{\mathrm{r}},
    \beta_{0},n)}{\tan\beta_{0}}
   =  
   \frac{
   \overline{K}_{\mathrm{r}}+n^2+\frac{\tan^2\beta_{0}}{2}
   }{
   \frac{1}{2}+\tan^2\beta_{0}
   } 
    \label{eq:RidgeModeCriterion}
\end{equation}
The left-hand side of this equation is the second term in the
right-hand side of equation~(\ref{eq:momentPreStress-Explicit}).  This
shows that the residual stress $M_{\II}^0$ is always positive at the
onset of bifurcation: \emph{the ridge buckling instability occurs in
the overcurved case}.

The reconstruction of the ridge mode is similar to that of the
centerline mode.  The ridge mode only involves bending $\Omega_{\II}$,
and the twist remains zero, $\Omega_{\III} = 0$: \emph{the ridge mode
is a pure bending mode}.  A consequence of this is that the centerline
remains planar.  The first two ridges modes ($n=2$ and $3$) are shown
in figure~\ref{fig:CenterlineModes}. By equation~(\ref{eq:RidgeModeCriterion}),
the first unstable ridge mode is the one with $n=2$ bumps.

This ridge mode has not been discussed earlier in the literature, to
the best of our knowledge.

\subsection{Interpretation of the buckling modes by a symmetry argument}

We have found two families of buckling modes: the centerline mode, and
the ridge mode.  Each buckling mode can occur with an arbitrary
azimuthal wavenumber, indexed by an integer $n\geq 2$.  The centerline
modes occur in pure twist: the twist $\Omega_{\III}$ is non-zero,
making the central ridge go out of plane, while the unconstrained
curvature $\Omega_{\II}$ remains unchanged, implying that the ridge
angle $\beta$ remains unperturbed.  By contrast, the ridge mode occurs
in pure bending: the twist $\Omega_{\III}$ remains zero, making the
central ridge remain planar, while the unconstrained curvature
$\Omega_{\II}$ is modulated together with the ridge angle $\beta$.

These features of the buckling modes can be interpreted based on
symmetry considerations.  In~\ref{sec:symmetry}, we identify a
symmetry of the equilibrium equations for the bistrip, which leaves
the circular base state invariant.  The two families of modes that we
have obtained are the eigenvectors of this symmetry.  Indeed, the
eigenvector corresponding to the eigenvalue $+1$ satisfies
$\tilde{\Omega}_{\III} = +\Omega_{\III}$ which, in view of
equation~(\ref{eq:TransformationOfStrainBySymmetry}) in the appendix,
implies $\Omega_{\III}=0$: this is the ridge mode.  The eigenvector
corresponding to the eigenvalue $-1$ satisfies $\tilde{\Omega}_{\II} =
-\Omega_{\II}$, implying $\Omega_{\II}=0$: this is the centerline
mode.  This symmetry explains why the eigenvalue problems for the
centerline and ridge modes in
equations~(\ref{eq:eigenvalueProblemCenterline})
and~(\ref{eq:strainIncrements-II-DiffEq}) are uncoupled, and why the
ridge angle $\beta$ is unaffected by the ridge mode, as observed in
previous work~\cite{Dias2012a}.

\section{Experiments}
\label{sec:experiments}

\subsection{Experimental buckling modes}

We confront the stability analysis carried out in the previous section
to experimental pictures of paper models.  An annular region is cut 
out in a piece of paper; as explained earlier in figure~\ref{fig:residual},
\begin{figure}
    \centering
    \includegraphics[width=14cm]{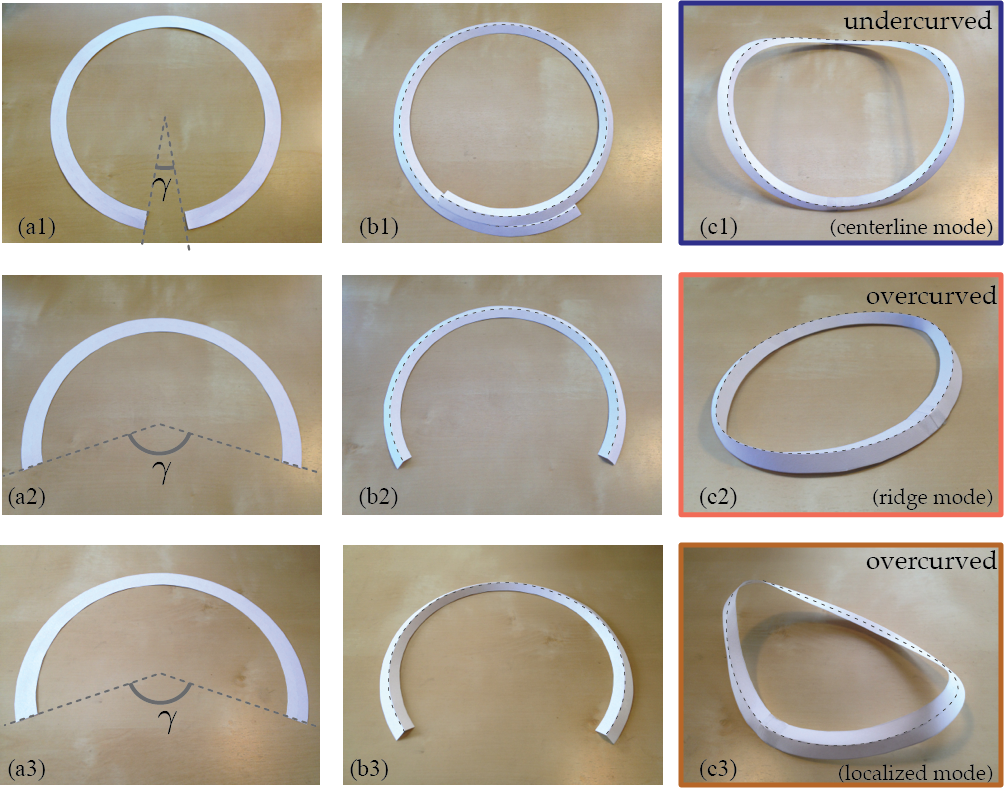}
    \caption{Observation of the centerline buckling mode (top row) and
    ridge buckling mode (bottom row) in a paper model.  Both modes
    have $n=2$ bumps, as predicted by the linear stability analysis.
    The bistrip is prepared as explained in figure~\ref{fig:residual}:
    (a) an annular region is cut out in a piece of paper and a sector
    of variable angular size $\gamma$ is removed from it; (b) the
    central ridge is pleated, leading to an increase in the curvature
    of the ridge, hence an overlap of the two free ends (undercurved
    case, b1) or a reduction of the gap between them (overcurved case,
    b2); (c) gluing the free ends together makes the bistrip buckle.
    In (b) and (c), the position of the ridge is highlighted by a
    dashed overlay.}
    \label{fig:OverUnderCurv}
\end{figure}
an angular sector of size $\gamma$ is removed, see (a) in
figure~\ref{fig:OverUnderCurv}, which sets the dihedral angle
$\beta_{0}$ of the circular solution by
equation~(\ref{eq:planar-beta0}); the permanent deformations involved
in pleating the ridge in step (b) amount to change the natural value
$\beta_{\mathrm{n}}$ of the ridge angle in the constitutive law.  The
circular configuration is not observed, as the bistrip buckles.  The
top row (a1--c1) in figure~\ref{fig:OverUnderCurv} corresponds to the
undercurved case: a buckling mode with $n=2$ bumps is observed, as
already reported in~\cite{Dias2012a}.  The features of the centerline
predicted by the linear stability analysis are confirmed: the
deformation involves twist, the centerline becomes non-planar and the
ridge angle remains uniform.

The second row (a2--c2) in figure~\ref{fig:OverUnderCurv} shows the
overcurved case, \emph{i.e.}\ when $\gamma$ is large enough and the
ends of the strip need to be \emph{pulled} to close up the bistrip.
The observed buckling mode is similar to the ridge mode predicted by
the linear stability analysis: the dihedral angle clearly varies along
the central fold in part (c2) of the figure, and the centerline
remains planar.  The observed mode corresponds to an azimuthal
wavevector $n=2$, as predicted by the theory.

Another buckling mode is observed in the experiments, which could not
be anticipated based on the linear stability analysis.  This mode,
shown in the bottom row (a3--c3) in figure~\ref{fig:OverUnderCurv} is
a non-planar pattern having a non-constant dihedral angle.  A striking
feature is that the deformation is localized at two opposite points,
where the curvature is quite large.  This mode is obtained for
slightly larger values of $\gamma$ than the ridge mode, \emph{i.e.}
for an even larger overcurvature.  This localized pattern is
essentially non-linear, and will be explained later on in
section~\ref{sec:AUTO}.

\begin{figure}
    \centering
    \includegraphics[width=10cm]{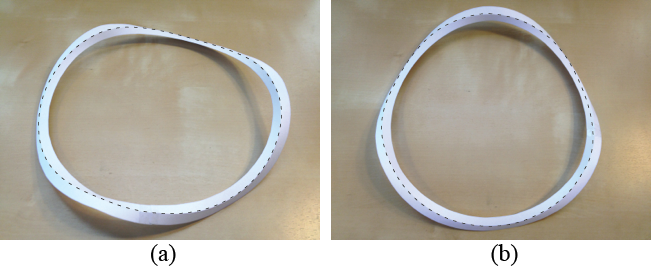}
    \caption{A higher-order centerline mode, with a wavenumber $n=3$,
    viewed from two angles.  This mode is achieved by compressing the
    natural mode $n=2$ between two plates.}
    \label{fig:HigherMode}
\end{figure}

In figure~\ref{fig:HigherMode}, we show that it is possible to force
the bistrip into a higher centerline mode, $n=3$.  Starting from the
natural buckling mode $n=2$, in figure~\ref{fig:OverUnderCurv}~(c1),
the higher mode can be obtained by squeezing the paper model between
two parallel plates.  When released, the shape with $n=3$ bumps
appears to be stable: it is likely to be a local equilibrium
configuration.

\subsection{Measuring the ridge stiffness}

\begin{figure}
    \centering
    \includegraphics[width=9cm]{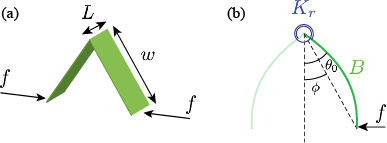}
    \caption{Sketch of the pinching experiment used to measure the
    dimensionless ridge stiffness $\overline{K}_r$.}
    \label{fig:MeasureKr}
\end{figure}

Here we show how the dimensionless ridge stiffness
$\overline{K}_{\mathrm{r}}$ can be measured experimentally.  The
value of $\overline{K}_{\mathrm{r}}$ is required to produce the
post-buckled solution in the following section.  The experimental set
up is depicted in figure~\ref{fig:MeasureKr}.
We cut out a short segment of the bistrip, with axial length $L=
1~\mathrm{cm}$.  The length $L$ and width $w=2~\mathrm{cm}$ are
comparable, and are much larger than the thickness $h\sim0.2~\mathrm{mm}$.  As sketched in the figure, a
pinching force $f$ is applied at the endpoints of the flaps.  By
measuring how much the flaps bend in response to this force, versus
how much the dihedral changes, one can find out the value of
$\overline{K}_{\mathrm{r}}$. 

To do so, we measured experimentally the values of the dihedral angle
$\theta_{0} = \frac{\pi}{2}-\beta$ and of the angle $\phi$ made by the
two endpoints (see figure) for various values of the applied force.
We simulated the problem of a 2D Elastica attached to an elastic hinge
numerically.  This problem depends only on the dimensionless stiffness
$\hat{K}_{\mathrm{r}} = \frac{w^2\,K_{\mathrm{r}}}{B}$.  We plotted
several parametric curves $f\mapsto
(\theta_{0}(\hat{K}_{\mathrm{r}},f),\phi(\hat{K}_{\mathrm{r}},f))$
corresponding to different values of $\hat{K}_{\mathrm{r}}$.  The
experimental datapoints were found to be distributed along one of the
simulation curves, and this allowed the parameter
$\hat{K}_{\mathrm{r}}$ to be determined.  This parameter was then
converted into the original dimensionless stiffness
$\overline{K}_{\mathrm{r}}$ defined in
equation~(\ref{eq:dimensionlessRidgeStiffness}) using the formula
$\overline{K}_{\mathrm{r}}= (r_{0}/w)^2\,\hat{K}_{\mathrm{r}}$.  For
the bistrips paper models used in the present paper, this yields
$\overline{K}_{\mathrm{r}} = 155$.

\section{Post-buckled solutions}
\label{sec:AUTO}

In this section, we investigate the post-buckled configurations of a
bistrip numerically, by solving the non-linear equations using a
continuation method.  The goal is to provide an example of application
of the bistrip model of section~\ref{sec:bistrip} in a fully
non-linear setting, to validate the assumptions and the predictions of
the linear stability analysis of section~\ref{sec:linearStability},
and to investigate the nature of the bifurcations.  We would also like
to explain the localized pattern observed in the experiments, which
the linear stability analysis could not reproduce.

The continuation method is implemented in two steps: the symbolic
calculation language
Mathematica~\cite{Wolfram-Research-Mathematica-Edition:-V9-2012} is
used to transform the equations for the strip in a set of first-order
differential equations, and export the right-hand sides as computer
code in the C language; in a second step, this code is used by the
continuation software
AUTO-07p~\cite{Doedel-Champneys-EtAl-AUTO-07p:-continuation-and-bifurcation-2007}
to produce the branches of equilibrium.

The unknowns are collected into a state vector $\underline{\mathcal{X}}(s)$,
\begin{equation}
    \underline{\mathcal{X}}(s) = \Big\{
    \underline{R}(s),
    \underline{D}_{\III}(s),
    \underline{D}_{\II}(s),
    \underline{N}(s),
    \beta(s),\beta'(s),\Omega_{\III}(s),
    \Lambda_{+}(s),\Lambda_{-}(s)
    \Big\}
    \textrm{,}
    \label{eq:auto-statevector}
\end{equation}
whose dimension is $N=17$.  The numerical continuation method requires
that we write the non-linear equations of equilibrium for the bistrip
in the form of $N$ first-order ordinary differential equations,
\begin{subequations}
    \label{eq:auto-BVP}
\begin{equation}
    \underline{\mathcal{X}}'(s) = \underline{\Phi}(
    \underline{\mathcal{X}}(s))
    \textrm{,}
    \label{eq:auto-differentialeq}
\end{equation}
together with $N$ boundary conditions,
\begin{equation}
    \underline{\Gamma}(\underline{\mathcal{X}}(0),
    \underline{\mathcal{X}}(L)) 
    =\underline{0}
    \textrm{.}
    \label{eq:auto-bcs}
\end{equation}
\end{subequations}

Let us now explain how the equilibrium equations for the bistrip are cast
in this form, starting with the differential
equation~(\ref{eq:auto-differentialeq}).  In terms of
$\underline{\mathcal{X}}(s)$, the following quantities are first
reconstructed: $\underline{D}_{I} = \underline{D}_{\II}\times
\underline{D}_{\III}$, $\Omega_{\II} = \kappa_{\mathrm{g}} /
\cos\beta$, $\underline{\Omega} =
\sum_{\mu=\II}^{\III}\Omega_{\mu}\,\underline{D}_{\mu}$.  Then, the
derivative of $\underline{\mathcal{X}}(s)$ is calculated as follows:
$\underline{R}' = \underline{D}_{\III}$, $\underline{D}_{\III}' =
\underline{\Omega}\times \underline{D}_{\III}$, $\underline{D}_{\II}'
= \underline{\Omega}\times \underline{D}_{\II}$, $\underline{N}' =
\underline{0}$; the derivative of $\beta$ is directly equated to the
following state variable $\beta'$; by inserting the full constitutive
law~(\ref{eq:effectiveConstitutiveLaw}) into the global balance of
moments~(\ref{eq:2Strips-BalanceOfMoments-Recap}) and the equilibrium
equation for the ridge~(\ref{eq:ridgeEquilibriumInCoordinates}), we
obtain four scalar equations, which we solve symbolically for
$\beta''$, $\Omega_{\III}'$, $\Lambda_{+}'$ and $\Lambda_{-}'$.  These
expressions for
$\{\underline{R}',\underline{D}_{\III}',\cdots,\Lambda_{-}'\}$ are
collected into a vector $\underline{\Phi}(
\underline{\mathcal{X}}(s))$ of length $N=17$, and the map
$\underline{\Phi}$ is implemented numerically in the C language.

The vector of the boundary conditions $\underline{\Gamma}$ is constructed
as follows.  We note that the solution is defined up to a rigid-body
motion, and remove this indeterminacy by the convention
$\underline{R}(\underline{0}) = \underline{0}$,
$\underline{D}_{\III}(0) - \underline{e}_{x} = \underline{0}$,
$\underline{D}_{\II}(0) - \underline{e}_{y} = \underline{0}$.  We also
enforce the periodicity conditions $\underline{R}(L) - \underline{R}(0)
= \underline{0}$, $\beta(L) - \beta(0) = 0$, $\beta'(L) - \beta'(0)= 0
$, $(\underline{D}_{\III})_{y}(L) - (\underline{D}_{\III})_{y}(0)= 0$,
$(\underline{D}_{\III})_{z}(L) - (\underline{D}_{\III})_{z}(0)= 0$,
$(\underline{D}_{\II})_{z}(L) - (\underline{D}_{\II})_{z}(0)= 0$.
This yields a total of $N=17$ scalar boundary conditions, which
are implemented as a map $\underline{\Gamma}(\underline{\mathcal{X}}(0),
\underline{\mathcal{X}}(L))$ in the C language.  It can be checked
that these periodicity conditions are necessary and sufficient to
warrant the periodicity of all the physical quantities of the strip,
such as $\Omega_{\mu}$, $\underline{D}_{\II}$, $\Lambda_{\pm}$, etc.

We work in a set of units such that $r_{0} = 1$, \emph{i.e.}\ the
curvilinear length of the ridge is $L=2\,\pi$, and the bending modulus
of the flaps is $B=1$.  The parameters of the simulation are the
natural angle $\beta_{\mathrm{n}}$ of the ridge, the ridge stiffness
$K_{\mathrm{r}}$ (which coincides with the rescaled one,
$\overline{K}_{\mathrm{r}}$, in this set of units), and the ridge
angle $\beta_{0}$ in the circular configuration.  The geodesic
curvature $\kappa_{\mathrm{g}} = \cos\beta_{0}$ is viewed as a
dependent variable (see equation~(\ref{eq:planar-beta0})).  We only
consider the fundamental buckling modes, $n=2$.

\begin{figure}[tbp]
    \centerline{\includegraphics[width=.99\textwidth]{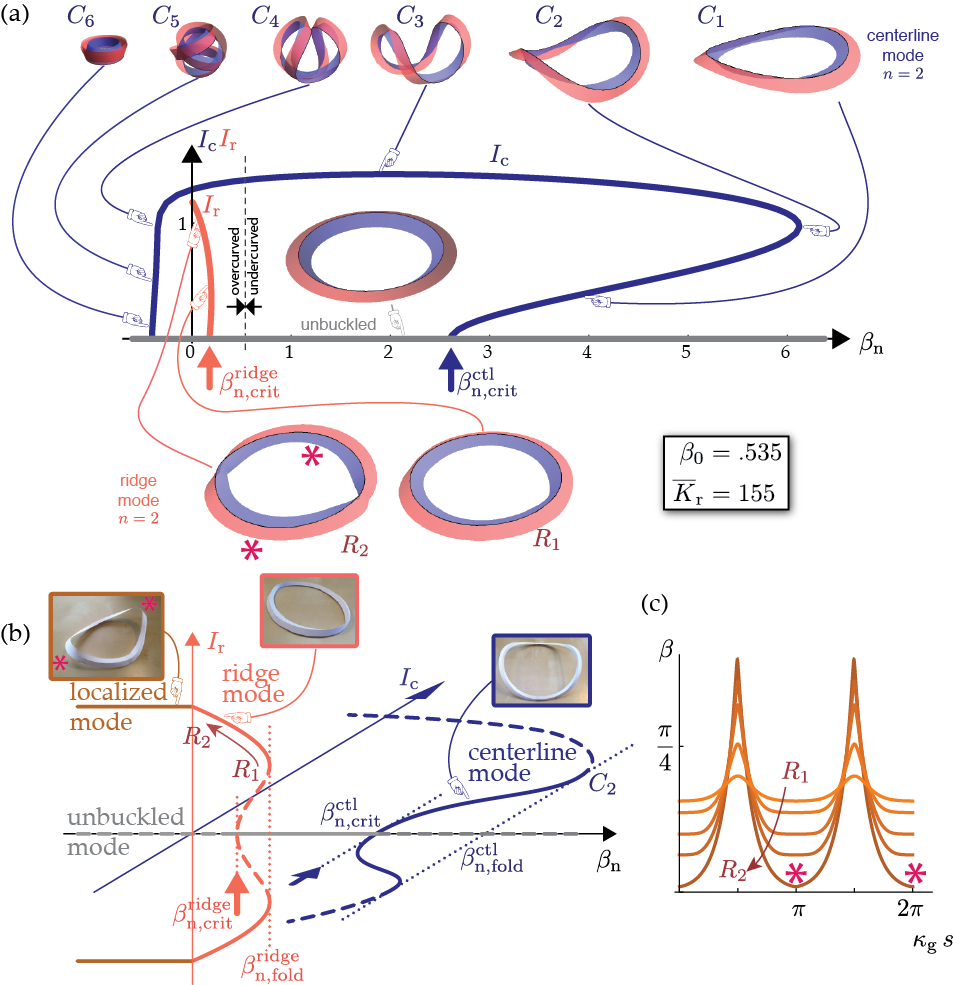}}
    \caption{Non-linear branches of equilibrium obtained by solving
    the fully non-linear equations of section~\ref{sec:Summary}.  (a)
    Results of the numerical continuation using AUTO for
    $\overline{K}_{\mathrm{r}} = 155$ and $\beta_{0} = .535$ (typical
    values of our experiments).  On the unbuckled branch (grey), both
    $I_{\mathrm{r}}$ and $I_{\mathrm{c}}$ cancel; on the ridge branch
    (orange), only $I_{\mathrm{r}}$ is non-zero; on the centerline
    branch (dark blue), only $I_{\mathrm{c}}$ is non-zero.  Only the
    buckled branches with azimuthal wavenumber ($n=2$) are shown.  The
    thick arrows below the $\beta_{\mathrm{n}}$ axis show the critical
    loads predicted by the linear stability analysis of
    section~\ref{sec:linearStability}.  (b) Stylized view of the same
    diagram representing the ridge and centerline modes in
    perpendicular directions, extending the branches to
    $I_{\mathrm{r}}<0$ and $I_{\mathrm{c}}<0$ by symmetry, stretching
    the ridge branch horizontally for better legibility, and showing
    the stable (solid curves) and unstable (dashed curves) portions of
    the branches.  (c) Plot of the dihedral angle $\beta$ as a
    function of the arclength $s$ for different solutions along the
    ridge branch: the dihedral angle progressively localizes as one
    moves towards the endpoint $R_{2}$ of the branch.  There, the
    bistrip flattens ($\beta=0$) at two opposite points (red
    asterisks).  Beyond this point $R_{2}$, for negative values of
    $\beta_{\mathrm{n}}$, the ridge branch connects to localized modes
    (brown curve in part b of the figure).}
    \protect\label{fig:PhaseDiagram}
\end{figure}

The boundary value problem in equations~(\ref{eq:auto-BVP}) is solved
using AUTO-07p.  A branch of solutions is produced by starting from
the circular configuration, with a radius $r_{0} = 1$ and a ridge
angle $\beta_{0}$.  The natural value of the ridge angle is
initialized to $\beta_{\mathrm{n}} = \beta_{0}$, and then used as a
continuation parameter: this mimics the act of creasing the central
fold further ($\beta_{\mathrm{n}} > \beta_{0}$), or flattening it
($\beta_{\mathrm{n}} < \beta_{0}$).  The equilibrium branches are
followed as $\beta_{\mathrm{n}}$ is varied.  Bifurcation diagrams
obtained in this way are shown in figure~\ref{fig:PhaseDiagram}.
The parameter $\beta_{0}$ and $\overline{K}_{\mathrm{r}}$ were set to
the values corresponding to our experiments.  In the diagram, we use
the buckling indicators $I_{\mathrm{c}}$ and $I_{\mathrm{r}}$ for the
centerline and for the ridge modes, respectively.  They are defined by
\begin{equation}
    I_{\mathrm{c}} = \left\langle 
    {\Omega_{\III}}^2\right\rangle^{1/2},\qquad
    I_{\mathrm{r}} = \left\langle {\beta'}^2\right\rangle^{1/2}
    \textrm{,}
    \label{eq:auto-buckling-indicators}
\end{equation}
where $\langle f \rangle = \frac{1}{L}\int_{0}^L f(s)\,\mathrm{d}s$ 
denotes the average of a function $f$.
These definitions are motivated by our stability
analysis: the centerline mode is a pure twist mode ($I_{\mathrm{c}}
\neq 0$ and $I_{\mathrm{r}}=0$), while the ridge mode is a pure
bending mode ($I_{\mathrm{c}} = 0$ and $I_{\mathrm{r}}\neq 0$). 

A stylized version of the numerical diagram obtained in
figure~\ref{fig:PhaseDiagram}(a) is shown in
figure~\ref{fig:PhaseDiagram}(b): in the latter, the ridge branch
(orange curve) has been stretched for better legibility.  First, we
note that the buckling thresholds predicted by the linear stability
analysis of section~\ref{sec:linearStability} are correct: they are
shown by the thick arrows below the $\beta_{\mathrm{n}}$ axis in part
(a) of the figure, and correspond exactly to the values of
$\beta_{\mathrm{n}}$ where the centerline or ridge branch meet the
unbuckled branch.  These buckling thresholds read, from
equations~(\ref{eq:centerlineModeCriterion})
and~(\ref{eq:RidgeModeCriterion}),
$\beta_{\mathrm{n,crit}}^\mathrm{ridge}(\overline{K}_{\mathrm{r}},
\beta_{0},n) = 0.177$ and
$\beta_{\mathrm{n,crit}}^\mathrm{ridge}(\overline{K}_{\mathrm{c}},
\beta_{0},n) = 2.612$, with $\overline{K}_{\mathrm{r}} = 155$,
$\beta_{0} = .535$ and $n=2$.  For reference, the transition from the
overcurved to the undercurved case, which can be found by solving
$M_{\II}^0=0$ for $Q_{0}$ in
equation~(\ref{eq:momentPreStress-Explicit}) and then for
$\beta_{\mathrm{n}}$ in the ridge's constitutive law, occurs at the
intermediate value $\beta_{\mathrm{n}} = 0.537$, as indicated by the
dashed vertical line in the figure (we noted earlier that the
centerline mode occurs in the undercurved case and the ridge mode in
the overcurved case).  The agreement of the non-linear post-buckling
and linear stability analyses on the initial thresholds confirms the
relevance of the simplified constitutive law
(\S\ref{ssec:nearCircularConstitutiveLaw}) to the linear stability
analysis.

Near $\beta_{\mathrm{n,crit}}^\mathrm{ridge}$, the centerline branch
emerges through a continuous pitchfork bifurcation.  The part of the
branch that extends between this initial bifurcation and the limit
point $C_{2}$ is stable, see part (b) of the figure.  This branch
spans the interval $\beta_{\mathrm{n,crit}}^\mathrm{ridge} \leq
\beta_{\textrm{n}} \leq \beta_{\mathrm{n,fold}}^\mathrm{ridge}$, where the
value of $\beta_{\mathrm{n}}$ corresponding to the limit point $C_{2}$
is $\beta_{\mathrm{n,fold}}^\mathrm{ridge} = 6.120$.  The presence of this
stable branch is consistent with the experimental observation of
saddle-like shapes for large enough undercurvature, as in the
experimental snapshot framed in dark blue in part (a) of the figure.
Note that the stability of the equilibria has been inferred by
comparing the elastic energy of the various types of solutions for a
given value of $\beta_{\mathrm{n}}$; a detailed analysis of stability
would be needed to confirm this.  Past the limit point $C_{2}$, the
centerline branch is unstable.  This unstable branch ultimately connects with
3-fold circular solutions, see snapshot $C_{6}$ in the figure.  We
ignore the self-contact that starts to take place beyond configuration
$C_{4}$.

Note that for the particular values $\overline{K}_{\mathrm{r}} = 155$
and $\beta_{0} = .535$ used to generate figure~\ref{fig:PhaseDiagram},
the stable centerline branch lies above
$\beta_{\mathrm{n,crit}}^\mathrm{ridge} = 2.612$.  This value is
beyond the maximum value $\beta_{\mathrm{n}} = \pi/2$ allowed by the
non-penetration condition of the flaps.  We conclude that for these
specific values of the parameters, the centerline buckling mode cannot
be observed.  The experimental snapshots of centerline modes shown in
figure~\ref{fig:PhaseDiagram}(b) and~\ref{fig:OverUnderCurv}(c1) were
indeed obtained for a much lower value of the angle $\beta_{0}$, and
the corresponding buckling thresholds
$\beta_{\mathrm{n,crit}}^\mathrm{ctl}$ predicted by
equation~(\ref{eq:centerlineModeCriterion}) are below $\pi/2$.

We now examine the ridge branch.  As emphasized in part (b) of the
figure, this branch is produced by a \emph{discontinuous} pitchofork
bifurcation: the weakly post-buckled ridge solutions exist for values
of $\beta_{\mathrm{n}}$ lying on the same side of the initial
threshold, $\beta_{\mathrm{n}} >
\beta_{\mathrm{n,crit}}^\mathrm{ridge}$, as the stable unbuckled
configuration.  This implies that the portion of the ridge branch
between the initial bifurcation threshold
$\beta_{\mathrm{n,crit}}^\mathrm{ridge}$ and the fold point
corresponding to the configuration $R_{1}$, which occurs for
$\beta_{\mathrm{n}} = \beta_{\mathrm{n,fold}}^\mathrm{ridge}= 0.196$,
is unstable: see dashed curve in part (b) of the figure.  Further
along the ridge branch, beyond the fold point $R_{1}$, it becomes
stable.  In the interval $\beta_{\mathrm{n,crit}}^\mathrm{ridge} = 0.177 <
\beta_{\mathrm{n}} < \beta_{\mathrm{n,fold}}^\mathrm{ridge} =
0.196$, both the ridge mode and the unbuckled solution are stable;
this interval has been stretched in figure~\ref{fig:PhaseDiagram}(b)
to improve legibility, but it is actually quite small.  The ridge mode
ceases to exist when $\beta_{\mathrm{n}}$ reaches a numerical value
equal to zero within numerical accuracy, which happens slightly beyond
the configuration labelled $R_{2}$ in the figure.  The plot of the
ridge angle $\beta(s)$ in figure~\ref{fig:PhaseDiagram}(c) shows that
the branch ends when the ridge angle, which progressively concentrates
into two narrow peaks, reaches the value $\beta=0$ at two opposite
points (asterisks).  The solutions that exist past this point cannot
be described with our equations, as we explicitly assumed $\beta\neq
0$ to derive equation~(\ref{eq:OmegaICancels}).  They can nevertheless
be discussed as follows.  When the ridge becomes flat, $\beta=0$, the
bistrip suddenly acquires another degree of freedom, which involves
bending both flaps into a cylindrical shape, with generatrices locally
perpendicular to the fold line.  This is exactly what happens in the
localized mode observed in the experiments, see the asterisks in the
experimental picture framed in brown in
figure~\ref{fig:PhaseDiagram}(b).  Therefore, we infer that the ridge
branch connects to a branch made of localized solutions in the region
$\beta_{\mathrm{n}}<0$, sketched by the brown line in part (b) of the
figure.  This is consistent with the experimental fact that planar
ridge solutions, which are rarely observed as they exist in a narrow
interval of $\beta_{\mathrm{n}}$, evolve into non-planar localized
solutions when the amount of overcurvature is increased.

Overall, the post-buckling diagram explains the three types of
patterns observed in the experiments.  The localized pattern could not
be anticipated by the linear stability analysis as it is produced by a
secondary bifurcation: along the ridge branch the dihedral angle
$\beta$ progressively concentrates until it reaches zero at two
opposite points, allowing the bistrip to become suddenly non-planar.

\section{Conclusion}

We have considered the large deformations of thin elastic strips,
whose width $w$ is much smaller than its length $L$ but much larger
than its thickness $h$: $h\ll w\ll L$.  For thin beams having a
slender cross-section, $h\ll w$, the classical rod theory of Kirchhoff
is known to be inapplicable.  Such beams are usually modeled using
Vlasov's theory for thin-walled beams.  Vlasov's models can be
justified from 3D elasticity but only in the case of moderate
deformations, when the cross-sections bend by a small amount.  In the
present work, however, we have considered large deformations of thin
strips.  The strip has been modeled as an inextensible plate, and the
geometric constraint of inextensibility has been treated exactly: the
cross-sections are allowed to bend by a significant amount.  Our model
extends the classical strip model of Sadowsky, and reformulate it in a
way that fits into the classical theory of rods.

To do this, we have identified the applicable geometrical constraints
and constitutive law.
The latter is non-linear because of underlying constraint of
developability.  The other classical equations for thin rods are
applicable (inextensibility and unshearability constraints, geometric
definition of the twist-curvature strain, equations of equilibrium).
Unifying the description of of elastic strips and rods allows the
large body of numerical and analytical methods developed for rods, to
be ported to strips: our stability analysis of the bistrip was adapted
from the classical stability analyses of elastic rings.

For the purpose of illustration, our model has been applied to a
specific geometry: the equilibria of a closed bistrip have been
analyzed.  Bifurcated solutions reported in prior work have been
interpreted based on a instability affecting the circular solutions.
Two other, novel types of patterns have been demonstrated in
experiments.  We have identified the residual bending moment in the
circular configuration as the stress driving these instabilities.  The
sign of this residual stress has been shown to determine which
buckling mode occurs.  A symmetry argument has been invoked to explain
the main features of these buckling modes.  The selection of the
wavenumber of the modes has been accounted for.  The non-linear
features of the instability have been explored numerically using a
continuation method.  In particular, we have identified a localized
mode, that can only be interpreted based on the post-buckling
analysis.


In future work, it would be interesting to extend the present approach
to corrugated shells.  Such shells are obtained by folding an elastic
plate along a family of folds that are locally parallel to each other.
The presence of the folds has a dramatic influence on the mechanical
behavior of the structure: for instance, they can make it behave like
a hyperbolic
shell, and can couple to two modes of bending~\cite{%
Norman20091624}.  These interesting behaviors
have awaken a marked interest recently.  So far, the analysis
of corrugated shells has been mainly carried out at the geometric level~\cite{%
Demaine-Demaine-EtAl-Nonexistence-of-pleated-folds:-2011,%
Seffen2012} or for specific fold geometries~\cite{Dias2012b,Wei2012}.
Generalizing the approach followed in our paper, it should be possible
to account for the presence of the folds through an effective
(homogenized) constitutive law.  This would make it possible to
bridge the gap between the literature on the mechanics of
elastic shells, and the young field of corrugated shells.

\appendix

\section{Constitutive law for an elastic rod with constraints}
\label{sec:VirtualWork}

In this appendix, we derive the general expression of the constitutive
law for an elastic rod subjected to kinematical constraint, and
possessing an internal degree of freedom.  It is considered
inextensible and the arclength is denoted by $s$.  A configuration of
the rod is parameterized by its centerline $\underline{r}(s)$, by an
orthonormal material frame $\underline{d}_{i}(s)$ with $i=1,2,3$, and
by an internal variable $k(s)$ (in the elastic strip model, this
internal variable is the transverse curvature $k_{11}$).  These
functions have to satisfy the Euler-Bernoulli and inextensibility
constraints in equation~(\ref{eq:tangent}--\ref{eq:orthonormal}).  The
Darboux vector $\underline{\omega}(s)$ is defined by
equation~(\ref{eq:darbouxVectorDefinition}), and we denote by
$\omega_{i} (s) = \underline{\omega}(s)\cdot \underline{d}_{i}(s)$ its
components in the material frame.  We assume that the elastic energy
$E_{\textrm{el}}$ of the rod is the integral of a density of elastic
energy, which is itself a function of the local twist and curvature
strains $(\omega_{1},\omega_{2},\omega_{3})$ and of the internal
parameter $k$.  In addition to the inextensibility and Euler-Bernoulli
constraints, the rod is subjected to a kinematical constraint which
writes $\mathcal{C}(\omega_{i}(s),k(s)) = 0$.

The rod is subjected to a density of external force $\underline{p}(s)$
and a density of external moment $\underline{q}(s)$ per arc-length
$\mathrm{d}s$.  Its equilibrium is governed by the principle of
virtual work.  The latter states that, for any virtual motion (see for
instance~\cite{Steigmann-Faulkner-Variational-theory-for-spatial-1993,%
Audoly-Pomeau-Elasticity-and-geometry:-from-2010}),
\begin{multline}
    -\int\left(\sum_{i=1}^3\frac{\delta 
    E_{\mathrm{el}}}{\delta\omega_i}\,\delta\omega_i+\frac{\delta 
    E_{\mathrm{el}}}{\delta k}\,\delta k\right)\mathrm{d}s
    +
    \int 
    \lambda\,\left(\sum_{i=1}^3\frac{\partial\mathcal{C}}{\partial\omega_i}\,\delta\omega_i+
    \frac{\partial\mathcal{C}}{\partial k}\,\delta k\right)\,\mathrm{d}s
    \cdots\\
    {}+\int \underline{n}\cdot (\delta 
    \underline{r}' - \delta \underline{d}_{3})\,\mathrm{d}s
    +\int(\underline{p}\cdot\delta \underline{r} + 
    \underline{q}\cdot\underline{\delta \theta})\,\mathrm{d}s
    = 0
    \textrm{,}
    \label{eq:pvv-plate}
\end{multline}
where $\delta E_{\mathrm{el}}/\delta \omega_{i}$ and $\delta
E_{\mathrm{el}}/\delta k$ denote the \emph{functional derivative} of
the elastic energy $E_{\mathrm{el}}$ with respect to $\omega_{i}(s)$
and $k(s)$, respectively.  In this equation, virtual (infinitesimal)
quantities are prefixed with the letter $\delta$: $\delta
\underline{r}$, $\delta \underline{d}_{3}$, $\delta
\underline{\theta}$, $\delta \omega_{i}$, $\delta k$ are the virtual
change of centerline, of tangent, the virtual infinitesimal rotation,
the virtual increment of twist and curvature, and of internal
parameter, respectively.  The first term in
equation~(\ref{eq:pvv-plate}) is the virtual internal work which
represents the elastic stress in the rod: as usual in the elastic
case, it is the opposite of the first variation of the elastic energy,
$-\delta E_{\mathrm{el}}$.  The second term takes into account
geometric constraint, $\mathcal{C}$ and $\lambda(s)$ is the associated
Lagrange multiplier.  The third term is associated with kinematic
constraint in equation~(\ref{eq:tangent}); the corresponding Lagrange
multiplier $\underline{n}(s)$ can be interpreted as the internal
force.  The last term is the virtual external work.  Note that in the
particular case where the external load is conservative, the principle
of virtual work~(\ref{eq:pvv-plate}) expresses the condition of
stationarity of the total energy subjected to the three kinematic
constraints listed above, as obtained by Lagrange's method of
constrained variations.

Cancelling the term proportional to $\delta k$ in 
equation~(\ref{eq:pvv-plate}), we obtain the condition of equilibrium 
with respect to the internal variable as
\begin{subequations}
    \label{eq:constitutiveLawForConstrainedRod}
    \begin{equation}
    -\frac{\delta 
    E_\mathrm{el}}{\delta k}
    + \lambda\,
    \frac{\partial\mathcal{C}}{\partial k} = 0
    \textrm{.}
    \label{eq:constitutiveLawForConstrainedRod-Constraint}
\end{equation}
Next, the strain increments are combined into a single vector defined
by $\delta\underline{\omega} =
\sum_{i=1}^3\,\delta\omega_{i}\,\underline{d}_{i}$, which can be
interpreted as the gradient of virtual rotation,
$\delta\underline{\omega} = \underline{\delta \theta}'$,
see~\cite{Audoly-Pomeau-Elasticity-and-geometry:-from-2010}.  We also
define
\begin{equation}
    \underline{m} = \sum_{i=1}^3\left(
    \frac{\delta 
    E_\mathrm{el}}{\delta\omega_i}-\lambda\,
    \frac{\partial\mathcal{C}}{\partial\omega_i}
    \right)\,\underline{d}_{i}
    \textrm{.}
    \label{eq:constitutiveLawForConstrainedRod-M}
\end{equation}
\end{subequations}
This allows us to rewrite the principle of virtual work in 
equation~(\ref{eq:pvv-plate}) as
\begin{equation}
    - \int \underline{m}\cdot \underline{\delta \theta}'
    \,\mathrm{d}s
    +\int \underline{n}\cdot (\delta 
    \underline{r}' - \underline{\delta \theta}\times \underline{r}')\,\mathrm{d}s
    +\int(\underline{p}\cdot\delta \underline{r} + 
    \underline{q}\cdot\underline{\delta \theta})\,\mathrm{d}s
    = 0
    \textrm{.}
    \label{eq:pvv-rod}
\end{equation}
This is the classical expression of the principle of virtual work for
inextensible Euler-Bernoulli rods \emph{without additional
constraints}.  By using geometrical identities and by integrating by
parts, one can 
show~\cite{Chouaieb-Kirchhoffs-problem-of-helical-solutions-of-uniform-2003,%
Steigmann-Faulkner-Variational-theory-for-spatial-1993,%
Audoly-Pomeau-Elasticity-and-geometry:-from-2010} that the
corresponding equations of equilibrium in strong form are the
Kirchhoff equations~(\ref{eq:classicalKirchhoffEquilibrium}).

We conclude than the equilibrium of a constrained elastic rod having
an internal degree of freedom $k(s)$ is governed by the equilibrium of
the internal degree of freedom in
equation~(\ref{eq:constitutiveLawForConstrainedRod-Constraint}) and by
the classical Kirchhoff equation for the equilibrium of rods.  In the
latter, one must use the expression of the internal moment
$\underline{m}(s)$ given by the constitutive
law~(\ref{eq:constitutiveLawForConstrainedRod-M}).
Equations~(\ref{eq:constitutiveLawForConstrainedRod-Constraint})
and~(\ref{eq:constitutiveLawForConstrainedRod-M}) are the main results
of this appendix.

%

\section{Symmetry relevant to the planar, circular state}
\label{sec:symmetry}

We identify a symmetry of the equations of equilibrium for a bistrip.
By this symmetry, the centerline gets reflected through to a plane.
The ridge does not get flipped, however: this symmetry is not merely a
pointwise reflection of the entire bistrip.  This symmetry accounts
for the two family of buckling modes from the circular configuration
(the centerline and ridge modes): this modes are the eigenvectors of
the symmetry operator.

\subsection{Definition of the symmetry}

Let us denote $\underline{\underline{G}}$ the reflection through the
$(xy)$ plane in the Euclidean space,
$\underline{\underline{G}}\cdot(x,y,z) = (x,y,-z)$.  We consider a
solution $\mathcal{S}$ of the equilibrium problem for the bistrip, as
summarized in section~\ref{sec:Summary}.  This solution is specified
by the functions
\begin{equation}
    \mathcal{S} = (\underline{R},\underline{D}_{I},\underline{D}_{\II},
    \underline{D}_{\III},
\beta, \underline{\Omega}, \Lambda_{+}, \Lambda_{-}, \underline{N},
\underline{M}, \underline{\Delta})
\textrm{.}
    \label{eq:a-solution}
\end{equation}

The symmetry is defined by its action onto the space of
configurations: it maps $\mathcal{S}$ onto another configuration
$\tilde{\mathcal{S}} =
(\tilde{\underline{R}},\tilde{\underline{D}}_{I},\cdots)$
defined by
\begin{subequations}
    \label{eq:symmetryDef}
    \begin{align}
        \tilde{\underline{R}}(s) & = \underline{\underline{G}}\cdot 
	\underline{R}(s)
        \label{eq:symmetryDef-R}\\
	\tilde{\underline{D}}_{\mu}(s) & = -\eta_{\mu}\,
	\underline{\underline{G}}\cdot \underline{D}_{\mu}(s)
	\label{eq:symmetryDef-Dj}\\
	\tilde{\beta}(s) & = \beta(s) 
	\label{eq:symmetryDef-beta} \\
	\tilde{\underline{\Omega}}(s) & =  
	-\underline{\underline{G}}\cdot\underline{\Omega}(s)
	\label{eq:symmetryDef-Omega} \\
	\tilde{\Lambda}_{\pm}(s) & = \pm\Lambda_{\pm}(s)
	\label{eq:symmetryDef-LambdaPM} \\
	\tilde{\underline{N}}(s) & = 
	\underline{\underline{G}}\cdot \underline{N}(s)
	\label{eq:symmetryDef-N} \\
	\tilde{\underline{M}}(s) & = 
	-\underline{\underline{G}}\cdot \underline{M}(s)
	\label{eq:symmetryDef-M} \\
	\tilde{\underline{\Delta}}(s) & = 
	\underline{\underline{G}}\cdot \underline{\Delta}(s),
	\label{eq:symmetryDef-Delta}
    \end{align}
\end{subequations}
where $\mu\in\{I,\II,\III\}$, and $\eta_{\mu}$ is the sign defined by
\begin{equation}
    \eta_{\mu} = (-1)^\mu = \begin{cases}
        -1 & \textrm{for $\mu=I,\III$} \\
        +1 & \textrm{for $\mu=\II.$} 
    \end{cases}
    \label{eq:eta}
\end{equation}

By equation \eqref{eq:symmetryDef-Dj}, the directors basis is mapped
to $\tilde{\underline{D}}_{I} = +\underline{\underline{G}}\cdot
\underline{D}_{I}$, $\tilde{\underline{D}}_{\II} =
-\underline{\underline{G}}\cdot \underline{D}_{\II}$ and
$\tilde{\underline{D}}_{\III} = +\underline{\underline{G}}\cdot
\underline{D}_{\III}$.  The minus sign in the definition of
$\tilde{\underline{D}}_{\II}$ preserves the right-handedness of the
frame.

The components of the Darboux vector in the local frame are
transformed according to $\tilde{\Omega}_{\mu} =
\tilde{\underline{\Omega}}\cdot \tilde{\underline{D}}_{\mu} =
(-\underline{\underline{G}}\cdot \underline{\Omega}) \cdot
(-\eta_{\mu}\,\underline{\underline{G}}\cdot \underline{D}_{\mu}) =
\eta_{\mu}\,\underline{\Omega}\cdot \underline{D}_{\mu} =
+\eta_{\mu}\,\Omega_{\mu}$.  This implies (\emph{o}) $\tilde{\Omega}_{I}=
\Omega_{I} = 0$: the symmetric configuration satisfies the
constraint~(\ref{eq:OmegaICancels-Recap}), and $(\emph{ii})$ the
unconstrained curvatures transform according to
\begin{subequations}
    \label{eq:TransformationOfStrainBySymmetry}
    \begin{align}
        \tilde{\Omega}_{\II} & = +\Omega_{\II}
        \label{eq:TransformationOfStrainBySymmetry-Curvature}\\
        \tilde{\Omega}_{\III} & = -\Omega_{\III}
	\textrm{.}
        \label{eq:TransformationOfStrainBySymmetry-Twist}
    \end{align}
\end{subequations}

By a similar argument, the internal moments $\underline{M}$and
$\underline{\Delta}$ are transformed according to $\tilde{M}_{j} =
+\eta_{j}\,M_{j}$ and $\tilde{\Delta}_{j} = -\eta_{j}\,\Delta_{j}$.
Using these transformation rules, it can be checked that the new state
$\tilde{\mathcal{S}}$ satisfies the equilibrium equations for a
bistrip summarized in section~\ref{sec:Summary}, when $\mathcal{S}$ is itself an equilibrium solution.  Note
that an external loading may break this symmetry, unless it is itself
symmetric --- in this paper, we ignore the external loading.

%

\section*{Acknowledgements}
MAD would like to thank financial support from NSF DMR 0846582 and NSF-supported MRSEC on Polymers at UMass (DMR-0820506) during the initial stages of this project.

\bibliographystyle{plain}


\end{document}